\documentclass[10pt,twocolumn,twoside] {IEEEtran}
\usepackage{graphicx,amsfonts}
\usepackage{subfig,color,framed}
\usepackage[bookmarks=false,hidelinks]{hyperref}
\usepackage{cite}
\ifCLASSINFOpdf
\else
\fi
\usepackage[cmex10]{empheq}
\usepackage{array,multirow}
\usepackage{mdwmath}
\usepackage{mdwtab}
\usepackage{algorithm}
\usepackage{varwidth}

\newtheorem{thm}{Theorem}
\newtheorem{lem}{Lemma}

\begin{document}
\setlength{\skip\footins}{2mm}%is the space between the text body and the footnotes:
%\addtolength{\footins}{-3mm}%is the space between the text body and the footnotes:
\addtolength{\abovedisplayskip}{-1.7mm}%space before maths
\addtolength{\belowdisplayskip}{-1.2mm}%space after maths
%\addtolength{\abovecaptionskip}{-2mm}%space above caption
%\addtolength{\belowcaptionskip}{-50mm}%space below caption
%\addtolength{\topmargin}{-10mm}%gap above header
%\addtolength{\topskip}{-2mm}%between header and text
%\addtolength{\parskip}{-2mm}%gap between paragraphs

\title{Speech Dereverberation Using Non-negative Convolutive Transfer Function and \\Spectro-temporal Modeling}

\author{Nasser~Mohammadiha, Simon~Doclo      % <-this % stops a space
\thanks{
\indent Nasser Mohammadiha has been with the Department of Medical Physics and Acoustics, University of Oldenburg, Germany, while working on this paper. He is currently with the Volvo Car Corporation, Sweden (e-mail: n.mohammadiha@gmail.com). Simon Doclo is with the Department of Medical Physics and Acoustics, University of Oldenburg, Germany (e-mail: simon.doclo@uni-oldenburg.de). This research was supported by the Cluster of Excellence 1077
"Hearing4all", funded by the German Research Foundation (DFG).}}% <-this % stops a space

\markboth{The paper is published at IEEE TRANSACTIONS ON AUDIO, SPEECH, AND LANGUAGE PROCESSING, 2016.}%
{Mohammadiha \MakeLowercase{\textit{et al.}}: Speech Dereverberation Using N-CTF}

\maketitle

%%%%%%%%%%%%%%%%%%%%%%%%%%%%%%%%%%%%%%%%%%%%%%%%%%%%%%%%%%%%%%%%%%%%%%%%%%%%%
\begin{abstract}
%%%%%%%%%%%%%%%%%%%%%%%%%%%%%%%%%%%%%%%%%%%%%%%%%%%%%%%%%%%%%%%%%%%%%%%%%%%%%
%\boldmath
This paper presents two single-channel speech dereverberation methods
to enhance the quality of speech signals that have been recorded in an enclosed space. For both methods, the room acoustics are modeled using a non-negative approximation of the convolutive transfer function (N-CTF), and to additionally exploit the spectral properties of the speech signal, such as the low-rank nature of the speech spectrogram, the speech spectrogram is modeled using non-negative matrix factorization (NMF). Two methods are described to combine the N-CTF and NMF models. In the first method, referred to as the \textit{integrated} method, a cost function is constructed by directly integrating the speech NMF model into the N-CTF model, while in the second method, referred
to as the \textit{weighted} method, the N-CTF and NMF based cost functions
are weighted and summed. Efficient update rules are derived to solve both optimization problems.
%, where the method of auxiliary function is utilized to derive efficient and novel iterative update rules.
In addition, an extension of the integrated method is presented, which exploits the temporal dependencies of the speech signal. Several experiments are
performed on reverberant speech signals with and without background noise, where the integrated method yields a considerably higher speech quality
than the baseline N-CTF method and a state-of-the-art spectral enhancement
method. Moreover, the experimental results indicate that the weighted method can even lead to a better performance in terms of instrumental quality measures, but that the optimal weighting parameter depends on the room acoustics and the utilized NMF model. Modeling the temporal dependencies in the integrated method was found to be useful only for highly reverberant conditions.
\end{abstract}

\begin{IEEEkeywords}
non-negative convolutive transfer function, spectral modeling, non-negative matrix factorization, speech dereverberation
\end{IEEEkeywords}

\IEEEpeerreviewmaketitle
%%%%%%%%%%%%%%%%%%%%%%%%%%%%%%%%%%%%%%%%%%%%%%%%%%%%%%%%%%%%%%%%%%%%%%%%%%%%%
\section{Introduction\label{sec:Introduction}}
%%%%%%%%%%%%%%%%%%%%%%%%%%%%%%%%%%%%%%%%%%%%%%%%%%%%%%%%%%%%%%%%%%%%%%%%%%%%%
Recording a sound source in an enclosed space with a microphone placed at a distance from the sound source typically results in a signal that is reverberant, i.e., affected by the acoustic reflections against walls and objects within the enclosure.
%is the phenomenon of sound persistence after it is produced in an enclosed space, which is caused by the reflections of the sound from the surrounding objects. Thus, when a distant microphone is used to record the audio signals in an enclosed space, the obtained signals are reverberant.
Reverberation may highly degrade the quality and
intelligibility of speech \cite{Naylor2010,Beutelmann2006}, and hence, in many speech communication applications,
such as hearing aids, hands-free telephony, and automatic
speech recognition, it is important to recover the (non-reverberant)
clean speech signal \cite{Naylor2010}.

Several methods have been developed for speech dereverberation,
i.e., for estimating the clean speech signal from the reverberant
microphone signals. These methods can be broadly classified into spectral enhancement methods, probabilistic model-based methods, and acoustic multichannel equalization methods.
%According to \cite{Naylor2010}, these methods can be classified into three main classes. First, beamforming methods where multiple microphone recordings are used to steer a beam in the direction of the desired source \cite{Doclo2003}.
Spectral enhancement methods \cite{Allen1977,Lebart2001,Habets2007,Habets2009,Kuklasinski2015,Schwarz2015} are usually
designed to only suppress the late reverberation and are typically based on a room acoustic model parameterized by the reverberation time $T_{60}$ and the direct-to-reverberation ratio (DRR).
These methods first estimate the power spectral density (PSD) of the late reverberation, and then use spectral enhancement, e.g., Wiener filtering,
to estimate the clean speech spectrogram.
Probabilistic model-based methods \cite{Nakatani2010,Jukic2014a,Schmid2014} typically use an autoregressive process to model the reverberation, where the clean speech spectral coefficients are modeled using a Gaussian or a Laplacian distribution. Speech dereverberation is then performed blindly by estimating the unknown model parameters, including the reverberation model parameters and the clean speech spectral coefficients.
%Third, blind system identification and inversion methods, which .
Finally, in acoustic multichannel equalization methods, the room impulse responses (RIR) between the source and the microphones are blindly estimated and used to design an equalization system \cite{Miyoshi1988,Huang2003,Lim2014}, where in theory perfect dereverberation can be achieved. However, multichannel equalization methods are only able to provide a good dereverberation performance if accurate estimates of the RIRs are available \cite{Kodrasi2013}. Although significant
progress has recently been made, designing effective and robust
speech dereverberation methods still remains a challenging task.
%In these methods, either the clean speech signal is directly estimated (treating the RIR as nuisance parameter), or the RIR is first estimated (treating the clean speech signal as nuisance parameter) and then the estimated RIR is inverted to estimate the clean speech signal.

In this paper, we focus on single-channel dereverberation in the magnitude spectrogram domain. We assume
that the magnitudes (or powers) of the short-time Fourier transform (STFT) coefficients
of the reverberant signal at each frequency bin are obtained by convolving
the STFT magnitudes (or powers) of the clean speech signal and the STFT magnitudes (or powers) of the RIR at that
frequency bin \cite{Kameoka2009}, which is referred to as the non-negative convolutive transfer function (N-CTF)
model hereafter. Although the N-CTF model only holds approximately, it can be advantageous as it does
not require to model the RIR phase, which is difficult to be robustly
modeled \cite{Kameoka2009}. Recently, speech dereverberation methods based on this model have been proposed \cite{Kameoka2009,Singh2010,Kumar2011a,Yu2012,Kallasjoki2014}, which simultaneously estimate the power spectrograms of the clean speech signal and the RIR, where a sparsity constraint is usually
imposed on each frequency bin of the speech spectrogram. Hence, since individual frequency bins are processed independently, these methods completely ignore the spectral structure of the speech signal.
The main contribution of this paper is to propose blind single-channel speech
dereverberation methods that jointly model the room acoustics using
the N-CTF model and the speech spectrogram using non-negative matrix factorization
(NMF).

NMF is a method to obtain a low-rank approximation of a non-negative
matrix \cite{Lee2000}. In speech processing, NMF is usually applied
on the speech magnitude (or power) spectrogram, where the spectrogram is approximated
by the product of two non-negative matrices, i.e., a basis matrix and an
activation matrix. The basis matrix represents the repeating spectral patterns, while the activation matrix represents the
presence of these patterns over time. As a result, it has been shown that NMF can be used
to efficiently exploit the low-rank nature of the speech spectrogram
and its dependency across the frequencies, and has been successfully applied
for different problems in speech processing, e.g., \cite{Cichocki2009,Smaragdis2006,Virtanen2007,Fevotte2009,Gemmeke2011a,Mohammadiha2013d}.

In this paper, we present two methods to combine the N-CTF-based acoustic model and the NMF-based spectral model, resulting in two different cost functions.
In the first method, referred to as the \textit{integrated} method, the speech NMF model is integrated into the N-CTF
model resulting in a combined cost function, while in the second method, referred to as the \textit{weighted} method, the NMF- and N-CTF-based cost functions are weighted and summed. By minimizing the obtained cost functions, we derive new
update rules to simultaneously estimate the spectrograms of the clean speech signal and the RIR. To additionally exploit the temporal dependencies of the speech signal (and hence spectro-temporal modeling of the speech signal) we use a frame-stacking method \cite{Gemmeke2011a}. The estimated speech spectrogram is then used to compute a real-valued spectral gain in order to estimate the clean speech signal from the reverberant signal. It should be mentioned that while the proposed weighted method in this paper is novel, some preliminary results for the integrated method and modeling temporal dependencies have been discussed in \cite{Mohammadiha2015}. In this paper, both dereverberation methods are compared with each other for several reverberant conditions, where we also investigate the dereverberation performance in the presence of background noise. For both dereverberation methods, the quality of the dereverberated signals is evaluated using three
instrumental measures. Experimental results show that by additionally
modeling the speech spectrogram using NMF in the N-CTF-based dereverberation, the instrumental speech
quality measures substantially improve compared to the baseline N-CTF-based method, and that the proposed speech dereverberation methods outperform a state-of-the-art dereverberation method based on spectral enhancement \cite{Habets2009}. Results indicate that the weighted method can lead to higher performance measures than the integrated method, but that the optimal weighting parameter between the NMF- and N-CTF-based cost functions depends on the room acoustics and the utilized NMF model. Finally, modeling the speech temporal dependencies
using a frame-stacking method was only found to be useful for highly reverberant
conditions when the low-rank NMF basis matrix was learned offline from
clean speech training data.

The paper is organized as follows. The N-CTF model and
its underlying assumptions are discussed in Section \ref{sec:N-CTF}.
In Section \ref{sec:Dereverberation-Based-on-NCTF} a single-channel dereverberation
method using the N-CTF model minimizing the generalized Kullback\textendash{}Leibler divergence is
reviewed. The dereverberation methods, combining the N-CTF and NMF
models, are presented in Section \ref{sec:proposed_methods} and their performance is experimentally evaluated in Section \ref{sec:Experimental-Results}.

%%%%%%%%%%%%%%%%%%%%%%%%%%%%%%%%%%%%%%%%%%%%%%%%%%%%%%%%%%%%%%%%%%%%%%%%%%%%%
\section{Non-negative Convolutive Transfer Function\label{sec:N-CTF}}
%%%%%%%%%%%%%%%%%%%%%%%%%%%%%%%%%%%%%%%%%%%%%%%%%%%%%%%%%%%%%%%%%%%%%%%%%%%%%
We consider an acoustic scenario, where a single speech source is recorded using one microphone in a reverberant enclosure without background noise\footnote{Please note that in Section V-B we will investigate the influence of background noise on the dereverberation performance.}. Let $s(n)$ and $h(n)$ denote the discrete-time clean speech signal
and the $M$-tap RIR between the speech source and the microphone, where $n$ denotes the time index. The reverberant
speech signal $y(n)$ is obtained by convolving $s(n)$ and $h(n)$, i.e.,
\begin{equation}
y\left(n\right)=\sum_{m=0}^{M-1}h\left(m\right)s\left(n-m\right).\label{eq:convolution-timeDomain}
\end{equation}

In the STFT domain, (\ref{eq:convolution-timeDomain}) can be equivalently
represented as \cite{Avargel2007}:
\begin{equation}
y_{c}\left(k,t\right)=\sum_{k^{\prime}=1}^{K}\sum_{\tau=0}^{L_{h}-1}h_{c}\left(k,k^{\prime},\tau\right)s_{c}\left(k^{\prime},t-\tau\right),\label{eq:stft_exact}
\end{equation}
where $y_{c}$, $s_{c}$, and $h_{c}$ denote the complex-valued STFT coefficients
of the microphone signal, the clean speech signal, and the RIR, $k$
and $k^{\prime}$ denote the frequency index, $K$ denotes the total number of frequency bins, $t$ denotes the frame
index, $t=1,\ldots T$, and $L_{h}$ denotes the RIR length in the STFT domain \cite{Avargel2007}. An approximation of (\ref{eq:stft_exact}), referred to as the convolutive
transfer function (CTF), has been proposed in \cite{Talmon2009}, where only
band-to-band filters, i.e., $k=k^{\prime}$, are used:
\begin{equation}
y_{c}\left(k,t\right)\approx\sum_{\tau=0}^{L_{h}-1}h_{c}\left(k,\tau\right)s_{c}\left(k,t-\tau\right),\label{eq:CTF-complex}
\end{equation}
where $h_{c}(k,k^{\prime},\tau)=h_{c}(k,k,\tau)$ has been replaced with $h_{c}\left(k,\tau\right)$
for simplicity. Based on (\ref{eq:CTF-complex}), it has been proposed
in \cite{Kameoka2009} to approximate the power spectrogram of the
reverberant signal as (see Appendix \ref{app: Appendix_NCTF_assumptions} for details):
\begin{equation}
\left|y_{c}\left(k,t\right)\right|^{2}\approx\sum_{\tau=0}^{L_{h}-1}\left|h_{c}\left(k,\tau\right)\right|^{2}\left|s_{c}\left(k,t-\tau\right)\right|^{2}.\label{eq:CTF-power}
\end{equation}
In this paper, we use a generalization of (\ref{eq:CTF-power}), i.e.
\begin{empheq}[box=\fbox]{equation}
y\left(k,t\right)\approx \hat{y}\left(k,t\right)= \sum_{\tau=0}^{L_{h}-1}h\left(k,\tau\right)s\left(k,t-\tau\right),\label{eq:N-CTF}
\end{empheq}
where $y(k,t)=\left|y_{c}\left(k,t\right)\right|^{p},$ with $p=1$ (magnitude spectrogram)
or $p=2$ (power  spectrogram),
and $s$ and $h$ are defined similarly as $y$. In (\ref{eq:N-CTF}), the spectrogram of the reverberant speech signal $y(k,t)$ is modeled as the convolution of the (non-negative) spectrogram of the clean speech signal $s(k,t)$ with non-negative coefficients $h(k,\tau)$ representing the acoustical environment, such that (\ref{eq:N-CTF}) is referred to as the non-negative convolutive transfer function (N-CTF) based acoustic model. Although (\ref{eq:N-CTF}) has been derived assuming $h({k,\tau})=|h_c(k,\tau)|^{p}$, it is important to note that in the remainder of the paper we will use the N-CTF model in (\ref{eq:N-CTF}) without necessarily relating the non-negative coefficients $h(k,\tau)$ to the time-domain RIR $h(n)$. In
matrix notation (\ref{eq:N-CTF}) can be written as:
\begin{equation}
\mathbf{Y}\approx\hat{\mathbf{Y}}=\mathbf{S}*\mathbf{H},\label{eq:N-CTF_matrix}
\end{equation}
where $\mathbf{Y}=[y(k,t)]$ is a $K\times T$-dimensional matrix, $\hat{\mathbf{Y}}=[\hat{y}(k,t)]$ is
the resulting approximation for $\mathbf{Y}$, the $K\times (T-L_h+1)$-dimensional matrix $\mathbf{S}$ and the $K\times L_h$-dimensional matrix $\mathbf{H}$ are defined similarly, and $*$ denotes a row-wise convolution of
two matrices.

\begin{figure}
\includegraphics[width=.995\columnwidth]{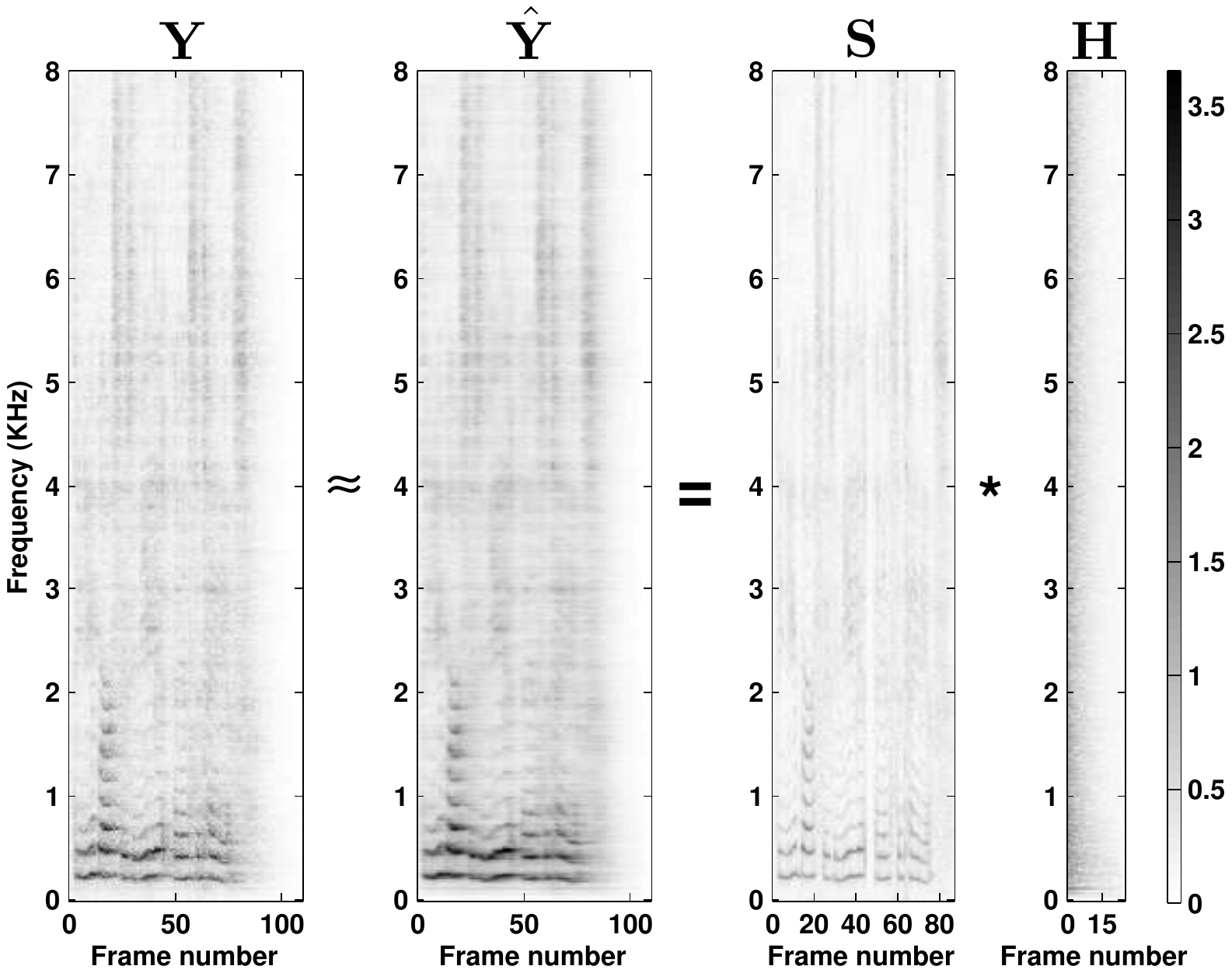}\caption{\label{fig:N-CTF-demonstration}N-CTF model approximation: The reverberant and clean speech magnitude spectrograms $\mathbf{Y}$ and $\mathbf{S}$ are obtained using  an STFT with
a frame length equal to 64 ms and 50\% overlapping frames. $\mathbf{H}$ is computed from a known RIR according
to \cite{Avargel2007}, where the off-diagonal elements have been set to zero.
The N-CTF model in (\ref{eq:N-CTF_matrix}) is used to approximate
$\mathbf{Y}$ as $\hat{\mathbf{Y}}=\mathbf{S}*\mathbf{H}$. }
\end{figure}

Fig. \ref{fig:N-CTF-demonstration} shows an example to visualize the quality of the approximation for the N-CTF model when using $h({k,\tau})=|h_c(k,\tau)|$ to approximate the reverberant magnitude spectrogram $\mathbf{Y}$ in (\ref{eq:N-CTF}), where a measured RIR ($T_{60}\approx680$ ms, direct-to-reverberation
ratio (DRR) around 0 dB) is used. The spectrogram $\mathbf{Y}$ is computed by
applying an STFT to $y(n)$ with a frame length equal to 64 ms and
50\% overlapping frames (with a sampling frequency equal to 16 kHz).
Similarly, the spectrogram $\mathbf{S}$ is computed by applying an STFT to the clean
speech signal $s(n)$. For the RIR, the representation from \cite{Avargel2007}
resulting in $h_{c}\left(k,k^{\prime},\tau\right)$ with $L_h=24$ is first obtained,
then the off-diagonal elements ($k\neq k^{\prime}$) are set to zero and $\mathbf{H}$ is computed as the magnitude of the resulting complex spectrogram. It has to be noted that computing $\mathbf{H}$ by setting the off-diagonal elements of $h_{c}\left(k,k^{\prime},\tau\right)$ to zero does not necessarily lead to the most accurate approximation of $\mathbf{Y}$ in \eqref{eq:N-CTF_matrix}. Nevertheless,
Fig.~\ref{fig:N-CTF-demonstration} shows that the N-CTF model, using the obtained $\mathbf{H}$,
results in a quite good, albeit smooth, approximation of the reverberant spectrogram.
In the following sections we will use the N-CTF model in (\ref{eq:N-CTF})
and blindly estimate both $\mathbf{H}$ and $\mathbf{S}$ to obtain the best
approximation of the reverberant spectrogram $\mathbf{Y}$.

%%%%%%%%%%%%%%%%%%%%%%%%%%%%%%%%%%%%%%%%%%%%%%%%%%%%%%%%%%%%%%%%%%%%%%%%%%%%%
\section{Review of Dereverberation Based on the N-CTF Model \label{sec:Dereverberation-Based-on-NCTF}}
%%%%%%%%%%%%%%%%%%%%%%%%%%%%%%%%%%%%%%%%%%%%%%%%%%%%%%%%%%%%%%%%%%%%%%%%%%%%%
The spectrogram of the clean speech signal $s(k,t)$ can be estimated
by minimizing a cost function measuring the approximation error between the reverberant speech spectrogram $y(k,t)$ and its approximation $\hat{y}(k,t)$ in (\ref{eq:N-CTF}). As a cost function we will use the generalized Kullback-Leibler (KL) divergence \cite{Lee2000}
between $y$ and $\tilde{y}$, which is a commonly used similarity measure to compare spectrograms, i.e.
\begin{equation}
Q=\sum_{k,t}KL\left(y\left(k,t\right)\left|\sum_{\tau=0}^{L_h-1}h\left(k,\tau\right)s\left(k,t-\tau\right)\right.\right),\label{eq:klDiv_n-ctf}
\end{equation}
where
\begin{align}
KL\hspace{-.4mm}\left(y\left(k,t\right)\left|\tilde{y}\left(k,t\right)\right.\right) & \hspace{-.4mm}=\hspace{-.4mm}y\left(k,t\right)\hspace{-.4mm}\log\frac{y\left(k,t\right)}{\tilde{y}\left(k,t\right)}\hspace{-.4mm}+\hspace{-.4mm}\tilde{y}\left(k,t\right)\hspace{-.4mm}-\hspace{-.4mm}y\left(k,t\right).\label{eq:kl-divergence}
\end{align}

The generalized KL divergence has been successfully applied for NMF-based speech enhancement and source separation  \cite{Virtanen2007,Gemmeke2011a,Mohammadiha2013g}, and has been used in \cite{Kallasjoki2014} for dereverberation.
In general, clean speech spectrograms can be assumed to be sparse, such that similarly to \cite{Kameoka2009}
it may be beneficial to add a sparsity-promoting term to \eqref{eq:klDiv_n-ctf}, obtaining the regularized cost function:
%{\small{
%\begin{empheq}[box=\fbox]{equation}
%Q=\sum_{k,t}KL\left(y\left(k,t\right)\left|\sum_{\tau=1}^{L_h}h\left(k,\tau\right)s\left(k,t-\tau\right)\right.\right)+\lambda\sum_{k,t}s\left(k,t\right),\label{eq:klDiv_n-ctf_sparse}
%\end{empheq}
%}}

{\small
\begin{equation}
Q=\sum_{k,t}KL\left(y\left(k,t\right)\left|\sum_{\tau=0}^{L_h-1}h\left(k,\tau\right)s\left(k,t-\tau\right)\right.\right)+\lambda\sum_{k,t}s\left(k,t\right),\label{eq:klDiv_n-ctf_sparse}
\end{equation}
}where $\lambda$ denotes the sparsity weighting parameter. Note that the cost function $Q$
does not include any criterion on the structure of the speech spectrogram
(except its sparsity), e.g., individual frequency bins are treated
independently.

Under the non-negativity constraints $h(k,\tau)\geq0$ and $s(k,t)\geq0$,
$s(k,t)$ and $h(k,\tau)$ can be estimated by minimizing the cost function in (\ref{eq:klDiv_n-ctf_sparse}). By applying the iterative learning method using auxiliary functions, which is briefly reviewed in Appendix \ref{app:iterative_learning}, the following iterative update rules can be obtained for $h(k,\tau)$ and $s(k,t)$. Here, $i$ denotes the iteration index, and $h^{i}(k,\tau)$ and $s^{i}(k,\tau)$
denote the estimates of $h(k,\tau)$ and $s(k,t)$ at the $i$-th iteration, respectively:
\begin{align}
h^{i+1}\left(k,\tau\right)=&h^{i}\left(k,\tau\right)\nonumber\times\\
&\frac{\sum_{t}y\left(k,t\right)s^{i}\left(k,t-\tau\right)/\tilde{y}\left(k,t\right)}{\sum_{t}s^{i}\left(k,t-\tau\right)},\label{eq:n-ctf based dereverb_h}
\end{align}

\begin{align}
s^{i+1}\left(k,t\right)=&s^{i}\left(k,t\right)\nonumber\times\\
&\frac{\sum_{\tau}y\left(k,t+\tau\right)h^{i+1}\left(k,\tau\right)/\tilde{y}\left(k,t+\tau\right)}{\sum_{\tau}h^{i+1}\left(k,\tau\right)+\lambda},\label{eq:n-ctf based dereverb_s}
\end{align}
where $\tilde{y}(k,t)=\sum_{\tau}h(k,\tau)s(k,t-\tau)$
is computed using the last available estimates of $h(k,\tau)$ and $s(k,t)$. To implement these update rules (and the ones in the next sections), a small positive number is typically added to the denominator to avoid division by zero. After convergence of the iterative learning method, the clean speech
signal $s(n)$ is estimated using the estimated spectrogram $s(k,t)$ and
inverse STFT, where the reverberant phase and the overlap-add
procedure are used.

%%%%%%%%%%%%%%%%%%%%%%%%%%%%%%%%%%%%%%%%%%%%%%%%%%%%%%%%%%%%%%%%%%%%%%%%%%%%%
\section{Dereverberation Based on Acoustic and Spectral Models \label{sec:proposed_methods}}
%%%%%%%%%%%%%%%%%%%%%%%%%%%%%%%%%%%%%%%%%%%%%%%%%%%%%%%%%%%%%%%%%%%%%%%%%%%%%
To exploit a-priori knowledge about the speech spectrogram, e.g., the low-rank nature of the spectrogram and its structure across frequencies, we propose to add a spectral model of the clean speech signal to the acoustic model in
(\ref{eq:N-CTF}). For this purpose, an NMF-based spectral model is introduced
in Section \ref{sub:NMF-based-Spectral-Model}. In the subsequent
sections, two methods are presented to combine the N-CTF-based acoustic
model in (\ref{eq:N-CTF}) with the NMF-based spectral model. These two methods exploit the
NMF model in a different way, resulting in different cost functions and update rules.
%,which are then minimized to estimate the spectrogram of the clean
%speech signal, i.e., $s(k,t)$, and the spectrogram of the RIR, i.e.,
%$h(k,t)$.
Section \ref{sub:Built-in-Method-to} presents the integrated method, where the NMF model is directly integrated
into the N-CTF model. In addition, an extension of the integrated method
is described in Section \ref{sub:Built-in-Method-to}
which exploits temporal dependencies using a frame-stacking method. Section \ref{sub:Weighted-Method-to} presents the weighted method, where the N-CTF- and NMF-based cost functions are weighted and summed.
%%%%%%%%%%%%%%%%%%%%%%%%%%%%%%%%%%%%%%%%%%%%%%%%%%%%%%%%%%%%%%%%%%%%%%%%%%%%%
\subsection{NMF-based Spectral Model\label{sub:NMF-based-Spectral-Model}}
%%%%%%%%%%%%%%%%%%%%%%%%%%%%%%%%%%%%%%%%%%%%%%%%%%%%%%%%%%%%%%%%%%%%%%%%%%%%%
Motivated by the successful modeling of speech spectrogram using non-negative
matrix factorization (NMF) in different applications, we propose to
use an NMF-based spectral model of the clean speech signal, i.e.,
\begin{empheq}[box=\fbox]{equation}
s\left(k,t\right)\approx\sum_{r=1}^{R}w\left(k,r\right)x\left(r,t\right),\label{eq:nmf-model}
\end{empheq}
where $w(k,r)$ and $x(r,t)$ are both non-negative, and $R$ denotes the number
of basis vectors in the $K\times R$-dimensional basis matrix $\mathbf{W}=[w(k,r)]$. In matrix
notation, (\ref{eq:nmf-model}) can be written as $\mathbf{S}\approx\mathbf{W}\mathbf{X}$,
where $\mathbf{S}=[s(k,t)]$ and $\mathbf{X}=[x(r,t)]$ denote the
speech spectrogram and the activation matrix, respectively. $R$
is typically chosen to be smaller than the dimensions of $\mathbf{S}$, so that
(\ref{eq:nmf-model}) results in a low-rank approximation of $\mathbf{S}$.

Given a speech spectrogram $\mathbf{S}$, the basis matrix $\mathbf{W}$
and the activation matrix $\mathbf{X}$ can be estimated by minimizing
a cost function measuring the distance between $\mathbf{S}$ and $\mathbf{W}\mathbf{X}$.
Common choices for the cost function are based on the generalized Kullback-Leibler (KL)
divergence, the Euclidean distance, or the Itakura\textendash{}Saito divergence.
The obtained cost functions usually correspond to different probabilistic
frameworks explaining how $\mathbf{S}$ is generated given $\mathbf{W}\mathbf{X}$.
For a given cost function, different optimization methods exist to iteratively estimate $\mathbf{W}$ and $\mathbf{X}$, where typically gradient-descent
update rules are applied for a number of iterations until a local minimum
of the cost function has been reached. Multiplicative update rules are popular methods for this purpose, which are obtained for a
particular choice of the step size in the gradient-descent update
rules \cite{Lee2000}.

%Fig. \ref{fig:Demonstration-of-NMF} shows an example where NMF minimizing
%the KL divergence is applied on a speech magnitude spectrogram $\mathbf{S}$,
%computed similarly to Section \ref{sec:N-CTF}, to obtain its low-rank
%approximation $\mathbf{\hat{S}}=\mathbf{W}\mathbf{X}$ with $R=25$.
%
%\begin{figure}
%\includegraphics[width=.995\columnwidth]{figures/nmf_demo}\caption{\label{fig:Demonstration-of-NMF}NMF demonstration: an STFT with frame
%length equal to 64 ms and 50\% overlapped frames (with a sampling
%frequency equal to 16 kHz) was used to obtain the speech magnitude
%spectrogram $\mathbf{S}\in\mathbb{R}^{513\times88}$. NMF was applied
%to estimate $\mathbf{W}$ (with $R=25$ basis vectors) and $\mathbf{X}$,
%and $\mathbf{S}$ is approximated as $\mathbf{S}\approx\mathbf{\hat{S}}=\mathbf{W}\mathbf{X}$.}
%\end{figure}

%%%%%%%%%%%%%%%%%%%%%%%%%%%%%%%%%%%%%%%%%%%%%%%%%%%%%%%%%%%%%%%%%%%%%%%%%%%%%
\subsection{Integrated Method to Combine N-CTF and NMF\label{sub:Built-in-Method-to}}
%%%%%%%%%%%%%%%%%%%%%%%%%%%%%%%%%%%%%%%%%%%%%%%%%%%%%%%%%%%%%%%%%%%%%%%%%%%%%
As the first method, we propose to directly integrate the NMF approximation of $s\left(k,t\right)$ in (\ref{eq:nmf-model}) into (\ref{eq:N-CTF}).
Consequently, the following cost function is obtained \cite{Mohammadiha2015}:
%the cost function in (\ref{eq:klDiv_n-ctf_sparse}) is equal to:%\hspace{-3mm}

\begin{framed}
{\small{
\begin{align}
\hspace{-4mm}L_{1}&=\sum_{k,t}KL\left(y\left(k,t\right)\left|\sum_{\tau=0}^{L_h-1}h\left(k,\tau\right)\sum_{r=1}^R w\left(k,r\right)x\left(r,t-\tau\right)\right.\right)\nonumber\\
&+\lambda\sum_{r,t}x\left(r,t\right),\label{eq:united_method}
\end{align}}}
\end{framed}
\noindent where the sparsity constraint is now imposed on the activations
$x$ since $s$ does not directly appear in (\ref{eq:united_method}).
This helps to obtain sparse estimates for $s$, considering the relation
between $s$ and $x$ in (\ref{eq:nmf-model}). The cost function
in (\ref{eq:klDiv_n-ctf_sparse}) is a special case of the cost function in (\ref{eq:united_method}) when the basis matrix $\mathbf{W}$ is a $K\times K$-dimensional
identity matrix. Moreover, the cost function utilized in \cite{Kallasjoki2014} is obtained as another special case, when $\mathbf{W}$ is a fixed matrix.

To minimize (\ref{eq:united_method}), the iterative learning method using auxiliary functions from Appendix \ref{app:iterative_learning} can be used, leading to the following multiplicative update
rules for $h$, $w$, and $x$:
\begin{equation}
h^{i+1}\left(k,\tau\right)=h^{i}\left(k,\tau\right)\frac{\sum_{t}y\left(k,t\right)\tilde{s}\left(k,t-\tau\right)/\tilde{y}\left(k,t\right)}{\sum_{t}\tilde{s}\left(k,t-\tau\right)},\label{eq:estimate_h_united}
\end{equation}

\begin{align}
w^{i+1}&\left(k,r\right)=w^{i}(k,r)\times\nonumber\\
&\frac{\sum_{t,\tau}y\left(k,t\right)h^{i+1}\left(k,\tau\right)x^{i}\left(r,t-\tau\right)/\tilde{y}\left(k,t\right)}{\sum_{t,\tau}h^{i+1}\left(k,\tau\right)x^{i}\left(r,t-\tau\right)},\label{eq:estimation_w_united}
\end{align}

\begin{align}
x^{i+1}&\left(r,t\right)=x^{i}\left(r,t\right)\times\nonumber\\
&\frac{\sum_{k,\tau}y\left(k,t+\tau\right)h^{i+1}\left(k,\tau\right)w^{i+1}\left(k,r\right)/\tilde{y}\left(k,t+\tau\right)}{\sum_{k,\tau}h^{i+1}\left(k,\tau\right)w^{i+1}\left(k,r\right)+\lambda},\label{eq:estimation_x_united}
\end{align}
where $\tilde{s}(k,t)=\sum_{r}w^{}(k,r)x^{}(r,t)$ and $\tilde{y}(k,t)=\sum_{\tau}h^{}(k,\tau)\tilde{s}(k,t-\tau)$ are computed using the last available estimates of the parameters. These update
rules can be efficiently implemented using the fast Fourier transform
(FFT) \cite{Kameoka2009}. To remove the scale ambiguity%
\footnote{Note that if $\hat{\mathbf{H}}$, $\hat{\mathbf{W}}$, and $\hat{\mathbf{X}}$
are a solution to (\ref{eq:united_method}), the same value
for $L_{1}$ can be obtained using $\alpha\hat{\mathbf{H}}$, $\hat{\mathbf{W}}/\alpha$,
and $\hat{\mathbf{X}}$ for all non-negative numbers $\alpha$.%
}, after each iteration each column of $\mathbf{W}$ is normalized
to sum to one, and the columns of $\mathbf{H}$ are element-wise divided
by the first column of $\mathbf{H}$ (resulting in an all-ones vector in the first
column of $\mathbf{H}$$)$. Moreover as suggested in \cite{Kallasjoki2014}, $h(k,\tau)$ is clamped to satisfy
$h(k,\tau)<h(k,\tau-1)$ for all $\tau$.

Let $\hat{\mathbf{W}}$, $\hat{\mathbf{X}}$,
and $\hat{\mathbf{H}}$ denote the obtained estimates
after convergence of the iterative method. One possible estimate
for the clean speech spectrogram $\mathbf{S}$ is given by
\begin{equation}
\hat{s}\left(k,t\right)=\sum_{r=1}^{R}\hat{w}\left(k,r\right)\hat{x}\left(r,t\right).\label{eq:estimate_s_nmf}
\end{equation}
Alternatively, the clean speech spectrogram can be estimated using
a time-varying gain function as $\hat{s}(k,t)=G(k,t)y(k,t)$, where
the gain function $G(k,t)$ is given by
\begin{equation}
G\left(k,t\right)=\frac{\sum_{r}\hat{w}\left(k,r\right)\hat{x}\left(r,t\right)}{\sum_{r,\tau}\hat{h}\left(k,\tau\right)\hat{w}\left(k,r\right)\hat{x}\left(r,t-\tau\right)},\label{eq:wiener_filtering}
\end{equation}
which was found to be particularly advantageous when the basis matrix $\mathbf{W}$ was learned offline from speaker-independent clean speech training data. Since \eqref{eq:estimate_s_nmf} directly uses the basis matrix to estimate the speech spectrogram, artifacts may be introduced, especially for unseen speakers. On the other hand, multiplying the reverberant spectrogram with the gain function in \eqref{eq:wiener_filtering} only uses the basis matrix in an indirect way with $\hat{w}(k,r)$ appearing both in the nominator and denominator, leading to less artifacts.
%This can be explained by noting that speaker-independent basis matrices may introduce some artifacts to represent a speech signal from an unseen speaker. Using Eq. \eqref{eq:wiener_filtering} can reduce such artifacts sine the basis matrix appears both in the nominator and denominator of the gain function, and hence the major artifacts cancel out during the gain calculation. However, Eq. \eqref{eq:estimate_s_nmf} directly uses the basis matrix to estimate the speech spectrogram and may introduce artifacts in the estimated speech signal.
Algorithm \ref{alg:Proposed-Built-in-Method} summarizes the integrated method.

%%%%%%%%%%%%%%%%%% algorithm %%%%%%%%%%%%%%%%%%%%%%%%%%%%%%%%%%%%%%%%%%%%%%
\begin{algorithm}[t]
Input : reverberant speech signal $y\left(n\right)$, output : dereverberated
speech signal $\hat{s}\left(n\right)$
\begin{enumerate}
\item Set the model parameters: $R$ (number of basis vectors), $It$ (number
of iterations), $\lambda$ (sparsity parameter), $L_{h}$ (RIR length
in the STFT domain), and $p$ (power).
\item Initialize $\mathbf{H}^{1}$, $\mathbf{W}^{1}$ and $\mathbf{X}^{1}$
with non-negative numbers (see Section \ref{sub:Evaluation_and_Implementation} for more details).
\item Compute the complex spectrogram $y_{c}\left(k,t\right)$ and the non-negative
spectrogram $y\left(k,t\right)=\left|y_{c}\left(k,t\right)\right|^{p}$
by applying an STFT to $y\left(n\right)$.
\item FOR $i=1$ to $It$
\begin{enumerate}
\item Compute $\mathbf{H}^{i+1}=[h^{i+1}(k,\tau)]$ using (\ref{eq:estimate_h_united})
\item Compute $\mathbf{W}^{i+1}=[w^{i+1}(k,r)]$ using (\ref{eq:estimation_w_united})
\item Compute $\mathbf{X}^{i+1}=[x^{i+1}(r,t)]$ using (\ref{eq:estimation_x_united})
\end{enumerate}
ENDFOR
\item Compute the time-varying gain function $G\left(k,t\right)$ using (\ref{eq:wiener_filtering}) and $h^{It}(k,\tau)$, $w^{It}(k,r)$, and $x^{It}(r,t)$.
\item Compute the dereverberated speech signal $\hat{s}\left(n\right)$
by applying an inverse STFT and the overlap-add procedure to $\hat{s}_c\left(k,t\right)=G\left(k,t\right)^{1/p}y_c(k,t)$.
\end{enumerate}
\caption{\label{alg:Proposed-Built-in-Method}Integrated method
to combine N-CTF and NMF}
\end{algorithm}

The presented integrated method
does not exploit temporal dependencies of the clean speech signal, i.e., consecutive frames are
processed independently. Temporal dependencies are however an important aspect of speech signals and have been shown
to be very beneficial for speech enhancement and source separation
\cite{Virtanen2007,Mohammadiha2013d,Mohammadiha2015a}. In order to investigate the benefit of modeling temporal dependencies for dereverberation, we propose an extension of the presented integrated method in Appendix \ref{app: Appendix_st} using the frame-stacking method \cite{Gemmeke2011a}.
In this method, a sliding window of size $T_{st}$ frames is used to divide the speech magnitude spectrogram into a number of overlapping windows. All the consecutive frames within each window are then stacked to obtain a high-dimensional vector, using which a high-dimensional matrix is constructed. Finally, NMF is applied to the obtained high-dimensional matrix to learn a high-dimensional basis matrix, which contains both spectral and temporal information of the speech signal. This frame-stacking method
is one of the important components in exemplar-based speech recognition \cite{Gemmeke2011a}. In Section \ref{sec:Experimental-Results}, the integrated methods with and without modeling temporal dependencies are evaluated and compared.
%%%%%%%%%%%%%%%%%%%%%%%%%%%%%%%%%%%%%%%%%%%%%%%%%%%%%%%%%%%%%%%%%%%%%%%%%%%%%
\subsection{Weighted Method to Combine N-CTF and NMF\label{sub:Weighted-Method-to}}
%%%%%%%%%%%%%%%%%%%%%%%%%%%%%%%%%%%%%%%%%%%%%%%%%%%%%%%%%%%%%%%%%%%%%%%%%%%%%
Since both the N-CTF acoustic model in (\ref{eq:N-CTF}) and the NMF
spectral model in (\ref{eq:nmf-model}) only hold approximately, it
may be beneficial to be able to give different weights to the corresponding cost
functions, which is not possible in the integrated method. This allows to give a larger weight to the more accurate approximation. For the N-CTF acoustic model in (\ref{eq:N-CTF}) we can use the cost function in (\ref{eq:klDiv_n-ctf_sparse}), whereas for the NMF spectral model in \eqref{eq:nmf-model} we can use a similar cost function based on generalized KL divergence, i.e.
\begin{equation}
P=\sum_{k,t}KL\left(s\left(k,t\right)\left|\sum_{r}w\left(k,r\right)x\left(r,t\right)\right.\right)+\lambda\sum_{r,t}x\left(r,t\right),\label{eq:nmf_klDiv}
\end{equation}
where we have additionally used a sparsity-promoting term to
encourage sparse estimates for the activations $x(r,t)$. The total cost function $L_{2}$ is now defined as the weighted sum
of $Q$ in (\ref{eq:klDiv_n-ctf_sparse}) and $P$ in (\ref{eq:nmf_klDiv}):
\begin{empheq}[box=\fbox]{equation}
L_{2}=\rho P+\left(1-\rho\right)Q,\label{eq:combined}
\end{empheq}
where $0<\rho<1$ denotes the weighting parameter.  Note that the same sparsity weighting parameter $\lambda$
is used in $P$ and $Q$ as both sparsity constraints used in $P$ and $Q$ are related to the sparsity of the speech spectrogram (cf. \eqref{eq:nmf-model}).

%we use an auxiliary function method to derive update rules to minimize $L_{2}$ w.r.t. $h$, $s$, $w,$ and $x$. In the following we first derive the update rules to minimize $L_{2}$ w.r.t. $s$ since it is a more involving problem; the minimization w.r.t. the other parameters can be solved similarly. Let estimates
%of $h$, $s$, $w$, and $x$ at the $i$-th iteration be denoted
In the following, we use the iterative learning method using auxiliary functions from Appendix \ref{app:iterative_learning} to derive update rules to minimize $L_{2}$ w.r.t. $h$, $s$, $w,$ and $x$. Here, we only derive the update rules for $s$ since the update rules for the other parameters can be derived similarly.
%Let estimates of $h$, $s$, $w$, and $x$ at the $i$-th iteration be denoted by $h^{i}$, $s^{i}$, $w^{i}$, and $x^{i}$, respectively.
Let $L_{2}(s)$ denote all terms depending on $s$ in (\ref{eq:combined}), where $h$, $w$, and $x$ are held fixed at $h^{i}$, $w^{i}$, and $x^{i}$, respectively:
\begin{align}
L_{2}\left(s\right)=&\rho\sum_{k,t}\left(s\left(k,t\right)\log\frac{s\left(k,t\right)}{\tilde{s}\left(k,t\right)}-s\left(k,t\right)\right)+\nonumber\\
&\left(1-\rho\right)\sum_{k,t}\left(\lambda s\left(k,t\right)+\sum_{\tau}h^{i}\left(k,\tau\right)s\left(k,t-\tau\right)\right)-\nonumber\\
&\left(1-\rho\right)\sum_{k,t}\left(y\left(k,t\right)\log\sum_{\tau}h^{i}\left(k,\tau\right)s\left(k,t-\tau\right)\right),\label{eq:definition_L2_s}
\end{align}
where $\tilde{s}(k,t)=\sum_{r}w^{i}(k,r)x^{i}(r,t)$. Minimizing $L_2(s)$ w.r.t. $s$ can be solved using Theorem \ref{thm:weighted_method}.
% $(i+1)$-th iteration
%given an estimate of all other parameters and $s^i$
%is derived in Theorem \ref{thm:weighted_method}.
%%%%%%%%%%%%%%%%%%%%%% Theorem %%%%%%%%%%%%%%%%%%%%%%%%
\begin{thm}
\label{thm:weighted_method}The function $L_{2}(s)$ in non-increasing
under the following update rule:
\begin{eqnarray}
s^{i+1}\left(k,t\right) & = & \frac{-c\left(k,t\right)}{\rho\mathcal{W}\left(-\frac{c\left(k,t\right)}{\rho}e^{b\left(k,t\right)/\rho}\right)},\label{eq:estimate_s_weighted}
\end{eqnarray}
where $\mathcal{W}$ is the Lambert W function \cite{Corless1996}, which is defined as:
\begin{equation}
z=\mathcal{W}\left(z\right)e^{{\mathcal{W}}\left(z\right)},\label{eq:lambert_W}
\end{equation}
and
\begin{align}
c\left(k,t\right)&=-\left(1-\rho\right)\sum_{\tau}\frac{y\left(k,t+\tau\right)s^{i}\left(k,t\right)h^{i}\left(k,\tau\right)}{\tilde{y}\left(k,t+\tau\right)},\label{eq:def_C}\\
b\left(k,t\right)&=\left(1-\rho\right)\left(\sum_{\tau}h^{i}\left(k,\tau\right)+\lambda\right)-\rho\log\tilde{s}(k,t),\label{eq:def_B}
\end{align}
where $\tilde{y}(k,t)=\sum_{\tau}h^{i}(k,\tau)s^{i}(k,t-\tau)$.\end{thm}
\begin{IEEEproof}
Since $-\log(x)$ is a convex function, using Lemma \ref{lem:jenens} from Appendix \ref{app:iterative_learning} with $x_{\tau}=h^{i}\left(k,\tau\right)s\left(k,t-\tau\right)$ and
$a_{\tau}=h^{i}\left(k,\tau\right)s^{i}\left(k,t-\tau\right)/\tilde{y}(k,t)$,
we can write:
\begin{align}
&-\log\sum_{\tau}h^{i}\left(k,\tau\right)s\left(k,t-\tau\right)\leq\nonumber\\
&-\sum_{\tau}\frac{h^{i}\left(k,\tau\right)s^{i}\left(k,t-\tau\right)}{\tilde{y}\left(k,t\right)}\log s\left(k,t-\tau\right)\nonumber\\
&-\sum_{\tau}\frac{h^{i}\left(k,\tau\right)s^{i}\left(k,t-\tau\right)}{\tilde{y}\left(k,t\right)}\log\frac{\tilde{y}(k,t)}{s^{i}(k,t-\tau)}.
\label{eq:Jensens_Neglog2}
\end{align}

Let us define the function $G(s,s^{i})$ as:
\begin{align}
&G\left(s,s^{i}\right)=\rho\sum_{k,t}\left(s\left(k,t\right)\log\frac{s\left(k,t\right)}{\tilde{s}\left(k,t\right)}-s\left(k,t\right)\right)+\nonumber\\
&\left(1-\rho\right)\sum_{k,t}\left(\lambda s\left(k,t\right)+\sum_{\tau}h^{i}\left(k,\tau\right)s\left(k,t-\tau\right)\right)-\nonumber\\
&\left(1-\rho\right)\sum_{k,t,\tau}\frac{y\left(k,t\right)h^{i}\left(k,\tau\right)s^{i}\left(k,t-\tau\right)}{\tilde{y}\left(k,t\right)}\log s\left(k,t-\tau\right)-\nonumber\\
&\left(1-\rho\right)\sum_{k,t,\tau}\frac{y\left(k,t\right)h^{i}\left(k,\tau\right)s^{i}\left(k,t-\tau\right)}{\tilde{y}\left(k,t\right)}\log\frac{\tilde{y}(k,t)}{s^{i}(k,t-\tau)}.
\end{align}

Using \eqref{eq:Jensens_Neglog2} it can be shown that $L_{2}(s)\leq G(s,s^{i})$. Since $G(s,s)=L_{2}(s)$, $G(s,s^{i})$ is an auxiliary function for $L_{2}(s)$.
%Since $-\log\sum_{\tau}h^{i}\left(k,\tau\right)s\left(k,t-\tau\right)$
%is convex, using Lemma \ref{lem:jenens} (with $s$ as the variable
%and using $a_{\tau}=h^{i}\left(k,\tau\right)s^{i}\left(k,t-\tau\right)/\tilde{y}(k,t)$)
%we can write:
%\begin{multline*}
%L_{2}\left(s\right)\leq G\left(s,s^{i}\right)=\\
%\rho\sum_{k,t}\left(s\left(k,t\right)\log\frac{s\left(k,t\right)}{\tilde{s}\left(k,t\right)}-s\left(k,t\right)\right)+\\
%\left(1-\rho\right)\sum_{k,t}\left(\lambda s\left(k,t\right)+\sum_{\tau}h^{i}\left(k,\tau\right)s\left(k,t-\tau\right)\right)-\\
%\left(1-\rho\right)\sum_{k,t,\tau}\frac{y\left(k,t\right)h^{i}\left(k,\tau\right)s^{i}\left(k,t-\tau\right)}{\tilde{y}\left(k,t\right)}\log s\left(k,t-\tau\right)-\\
%\left(1-\rho\right)\sum_{k,t,\tau}\frac{y\left(k,t\right)h^{i}\left(k,\tau\right)s^{i}\left(k,t-\tau\right)}{\tilde{y}\left(k,t\right)}\log\frac{\tilde{y}(k,t)}{s^{i}(k,t-\tau)}.
%\end{multline*}
Differentiating $G(s,s^{i})$ w.r.t. $s(k,t)$ and rearranging the
terms we obtain:
%\footnote{To avoid confusion with indices, it is easier to differentiate $G(s,s^{i})$
%w.r.t. $s(k^{\prime},t^{\prime})$, and to obtain a new estimate for
%$s(k^{\prime},t^{\prime})$. We use $k$ and $t$ in the presentation
%for better readability.}:
\begin{equation}
\frac{\partial G(s,s^{i})}{\partial s\left(k,t\right)}=\frac{c\left(k,t\right)}{s\left(k,t\right)}+\rho\log s\left(k,t\right)+b\left(k,t\right),\label{eq:derivative}
\end{equation}
where $c(k,t)$ and $b(k,t)$ are defined in (\ref{eq:def_C}) and
(\ref{eq:def_B}), respectively. Defining $z(k,t)=-c(k,t)/(\rho s(k,t))$
and setting (\ref{eq:derivative}) equal to zero we obtain:

\begin{equation}
z\left(k,t\right)+\log z\left(k,t\right)=\log\left(-\frac{c\left(k,t\right)}{\rho}e^{b\left(k,t\right)/\rho}\right).\label{eq:Lambert}
\end{equation}
The solution to (\ref{eq:Lambert}) is given by
\begin{equation}
z\left(k,t\right)=\mathcal{W}\left(-\frac{c\left(k,t\right)}{\rho}e^{b\left(k,t\right)/\rho}\right).\label{eq:solution_z}
\end{equation}
Substituting $z(k,t)$ with $-c(k,t)/(\rho s^{i+1}(k,t))$ leads to
(\ref{eq:estimate_s_weighted}). The theorem is now proved using Lemma
\ref{lem:auxiliary} from Appendix \ref{app:iterative_learning}.
\end{IEEEproof}

In a similar way, the multiplicative update rules for $w$, $x$, and $h$ can be derived as:
\begin{align}
h^{i+1}\left(k,\tau\right)=&h^{i}\left(k,\tau\right)\times \nonumber\\
&\frac{\sum_{t}y\left(k,t\right)s^{i+1}\left(k,t-\tau\right)/\tilde{y}\left(k,t\right)}{\sum_{t}s^{i+1}\left(k,t-\tau\right)},\label{eq:update_h_weighted}
\end{align}

\begin{align}
w^{i+1}(k,r)=&w^{i}(k,r)\times\nonumber\\
&\frac{\sum_{t}s^{i+1}\left(k,t\right)x^{i}\left(r,t\right)/\tilde{s}\left(k,t\right)}{\sum_{t}x^{i}\left(r,t\right)},\label{eq:update_w_weighted}
\end{align}

\begin{align}
x^{i+1}\left(r,t\right)=&x^{i}\left(r,t\right)\times\nonumber\\
&\frac{\sum_{k}s^{i+1}\left(k,t\right)w^{i+1}\left(k,r\right)/\tilde{s}\left(k,t\right)}{\sum_{k}w^{i+1}\left(k,r\right)+\lambda},\label{eq:update_x_weighted}
\end{align}
where $\tilde{s}(k,t)$ and $\tilde{y}\left(k,t\right)$, defined after (\ref{eq:definition_L2_s}) and (\ref{eq:def_B}), are computed using the last available estimates of the parameters. Algorithm
\ref{alg:Proposed-Weighted-method} summarizes the proposed weighted method.

%%%%%%%%%%%%%%%%%%%%%%%%% algorithm %%%%%%%%%%%%%%%%%%%%%%%%%%%
\begin{algorithm}[t]
Input : reverberant speech signal $y\left(n\right)$, output : dereverberated
speech signal $\hat{s}\left(n\right)$
\begin{enumerate}
\item Set the model parameters: $R$ (number of basis vectors), $It$ (number
of iterations), $\lambda$ (sparsity parameter), $L_{h}$ (RIR length
in the STFT domain), $p$ (power), and
$\rho$ (weighting parameter).
\item Initialize $\mathbf{H}^{1}$, $\mathbf{S}^{1}$, $\mathbf{W}^{1}$
and $\mathbf{X}^{1}$ with non-negative numbers (see Section \ref{sub:Evaluation_and_Implementation} for more details).
\item Compute the complex spectrogram $y_{c}\left(k,t\right)$ and the non-negative
spectrogram $y\left(k,t\right)=\left|y_{c}\left(k,t\right)\right|^{p}$
by applying an STFT to $y\left(n\right)$.
\item FOR $i=1$ to $It$
\begin{enumerate}
\item Compute $\mathbf{H}^{i+1}=[h^{i+1}(k,\tau)]$ using (\ref{eq:update_h_weighted})
\item Compute $\mathbf{S}^{i+1}=[s^{i+1}(k,t)]$ using (\ref{eq:estimate_s_weighted})
\item Compute $\mathbf{W}^{i+1}=[w^{i+1}(k,r)]$ using (\ref{eq:update_w_weighted})
\item Compute $\mathbf{X}^{i+1}=[x^{i+1}(r,t)]$ using (\ref{eq:update_x_weighted})
\end{enumerate}
ENDFOR
\item Compute the time-varying gain function as
\begin{equation}
G\left(k,t\right)=\frac{s^{It}\left(k,t\right)}{\sum_{\tau=0}^{L_{h}-1}h^{It}\left(k,\tau\right)s^{It}\left(k,t-\tau\right)}.
\label{eq:gain_weighted}
\end{equation}
\item Compute the dereverberated speech signal $\hat{s}\left(n\right)$
by applying an inverse STFT and the overlap-add procedure to $\hat{s}_c\left(k,t\right)=G\left(k,t\right)^{1/p}y_c(k,t)$.
\end{enumerate}
\caption{\label{alg:Proposed-Weighted-method}Weighted method
to combine N-CTF and NMF}
\end{algorithm}

%%%%%%%%%%%%%%%%%%%%%%%%%%%%%%%%%%%%%%%%%%%%%%%%%%%%%%%%%%%%%%%%%%%%%%%%%%%%%
\section{Experimental Results \label{sec:Experimental-Results}}
%%%%%%%%%%%%%%%%%%%%%%%%%%%%%%%%%%%%%%%%%%%%%%%%%%%%%%%%%%%%%%%%%%%%%%%%%%%%%
This section presents the evaluation setup and the results of several experiments to evaluate the performance of the
proposed single-channel speech dereverberation methods. Section \ref{sub:Evaluation_and_Implementation}
describes the acoustic setup, the used performance measures and the implementation
details of the proposed methods, e.g., parameter values and initialization. Experimental
results for the integrated method with and without temporal modeling are presented
in Section \ref{sub:results_integrated_method}, where the performance
of the integrated method is also compared to the performance of a state-of-the-art
spectral enhancement method and the dereverberation performance in the presence of background noise is investigated. Section \ref{sub:Weighted-Method-to}
compares the performance of the integrated and weighted
methods. Experimental results to analyze the effect of the parameters
on the performance of the developed methods are described in Section
\ref{sub:Important-Parameters}.

%%%%%%%%%%%%%%%%%%%%%%%%%%%%%%%%%%%%%%%%%%%%%%%%%%%%%%%%%%%%%%%%%%%%%%%%%%%%%
\subsection{Evaluation Setup and Implementation Details\label{sub:Evaluation_and_Implementation} }
%%%%%%%%%%%%%%%%%%%%%%%%%%%%%%%%%%%%%%%%%%%%%%%%%%%%%%%%%%%%%%%%%%%%%%%%%%%%%
To evaluate the performance of the dereverberation methods for a wide variety of acoustic conditions, the reverberant microphone signals were generated by convolving clean speech signals with three different measured RIRs with reverberation
times $T_{60}\approx430$ ms, $T_{60}\approx680$ ms, and $T_{60}\approx640$
ms, and direct-to-reverberation ratios (DRR) around 5, 0, and 12 dB,
respectively. As clean speech signals, 16 different sentences (uttered by 16 speakers) from the TIMIT database
\cite{Garofolo1993} were used to make the results independent of the speech
material. The sampling frequency was 16 kHz and the STFT frame length
and overlap length were equal to 64 ms and 32 ms, respectively, where
a square-root Hann window was used both for the STFT analysis and synthesis.

The dereverberation performance is evaluated using the following instrumental measures: the perceptual evaluation of speech quality (PESQ) \cite{PESQ2000}
and the cepstral distance (CD) \cite{Hu2008}, both using the clean speech signal
as the reference signal, and the reverberation decay tail (RDT) \cite{Wen2006}. The RDT is defined as the ratio of the amplitude and decay rate of the
exponential curves, which are fitted to the bark spectral difference between
the reverberant and the clean signal, normalized to the amplitude of
the direct component. A higher value for PESQ and a lower value for
CD and RDT indicate a better performance. These measures exhibit a high correlation with subjective listening tests evaluating the quality of the dereverberated speech signals \cite{Goetze2014}. The \textit{improvements}
obtained in PESQ, denoted by $\Delta\text{PESQ}$, and the
\textit{reductions} obtained in CD and RDT, denoted by $\Delta\text{CD}$
and $\Delta\text{RDT}$, respectively, for the considered dereverberation methods are shown in the subsequent
sections for better readability. To compute these  differential scores, the score of the dereverberated signal is compared to the score of the reverberant microphone signal.

The proposed integrated and weighted methods are
applied on the magnitude spectrogram of the reverberant microphone signals, i.e.,
$p=1$, since using the magnitude spectrogram
resulted in a better dereverberation performance compared to the power
spectrogram (cf. Section \ref{sub:Important-Parameters}). For both
integrated and weighted methods, two different ways were used to learn the basis matrix
$\mathbf{W}$:
\begin{itemize}
  \item The basis matrix $\mathbf{W}$, with $R=100$ columns, was estimated from the
reverberant speech signal. These variants are referred to as N-CTF+NMF and
N-CTF+NMF (w), respectively, where the suffix '(w)' is used throughout
the experiments to indicate the weighted method.
  \item The basis matrix $\mathbf{W}$ was learned offline from clean speech training data, consisting of 250 sentences uttered by 27 speakers, disjoint from
the test data, and was held fixed in the experiments. We consider two types of speaker-independent NMF
models: 1) a low-rank NMF model with $R=100$ (N-CTF+NMF+LR), and 2) an overcomplete NMF model with $R=3000$ (N-CTF+NMF+OC). The basis matrix of this overcomplete model was constructed
by sampling from the magnitude spectrogram of the training data using
a uniform random walk method \cite{Mohammadiha2014}.
\end{itemize}

The RIR length in the STFT domain $L_{h}$ was set to 10, corresponding
to 320 ms, independent of the reverberation time. Each row of $\mathbf{H}$
was initialized identically using a linearly-decaying envelope, while
$\mathbf{W}$ and $\mathbf{X}$ were first initialized by random non-negative numbers and then updated by iterating the
standard NMF update rules on the spectrogram of the reverberant signal $\mathbf{Y}$ for 10 times. For the weighted
method, $\mathbf{S}$ was initialized with $\mathbf{Y}$. The $T_{st}$ parameter in the frame-stacking method to model the temporal dependencies was set to 6. For both methods, the sparsity parameter $\lambda$ was set to $\frac{0.1}{KT}\sum_{k,t}y(k,t)$ and to encourage sparser solutions, after each iteration
the estimates of $x$ and $s$ were raised
to a power $\phi_{x}$ as proposed in \cite{Cichocki2006}.
%, where we experimentally set $\phi_{x}=1.02$ when $R=100,$ and $\phi_{x}=1.05$ when $R=3000$.
The maximum number of iterations was experimentally set to 20 for the integrated method and 70 for the weighted method.

The proposed methods are compared to the baseline N-CTF method (cf. Section
\ref{sec:Dereverberation-Based-on-NCTF}), which does not use any spectral model, and a state-of-the-art speech spectral
enhancement method, where the late reverberant spectral variance
was estimated using \cite{Habets2009} based on \textit{oracle} $T_{60}$ and DRR values computed from the RIR, and the log-spectral amplitude
estimator \cite{Ephraim1985} was used to estimate the clean speech STFT coefficients; this method is referred to as the SE method in the following. It should be noted that the N-CTF-based dereverberation methods including the proposed methods are batch methods, i.e., the whole reverberant microphone signal corresponding to a short utterance is used to estimate the clean speech signal, while the SE method using \cite{Habets2009} is an on-line method. Developing an on-line N-CTF-based dereverberation method, similar to on-line NMF-based speech enhancement such as \cite{Mohammadiha2013d} remains a question for future research.
%%%%%%%%%%%%%%%%%%%%%%%%%%%%%%%%%%%%%%%%%%%%%%%%%%%%%%%%%%%%%%%%%%%%%%%%%%%%%
\subsection{Integrated Method\label{sub:results_integrated_method}}
%%%%%%%%%%%%%%%%%%%%%%%%%%%%%%%%%%%%%%%%%%%%%%%%%%%%%%%%%%%%%%%%%%%%%%%%%%%%%
%\addtocounter{subfigure}{3}
\begin{figure}
\centering
\subfloat[\label{fig:Reverberant-speech-signal}Reverberant speech signal]{\includegraphics[width=.995\columnwidth]{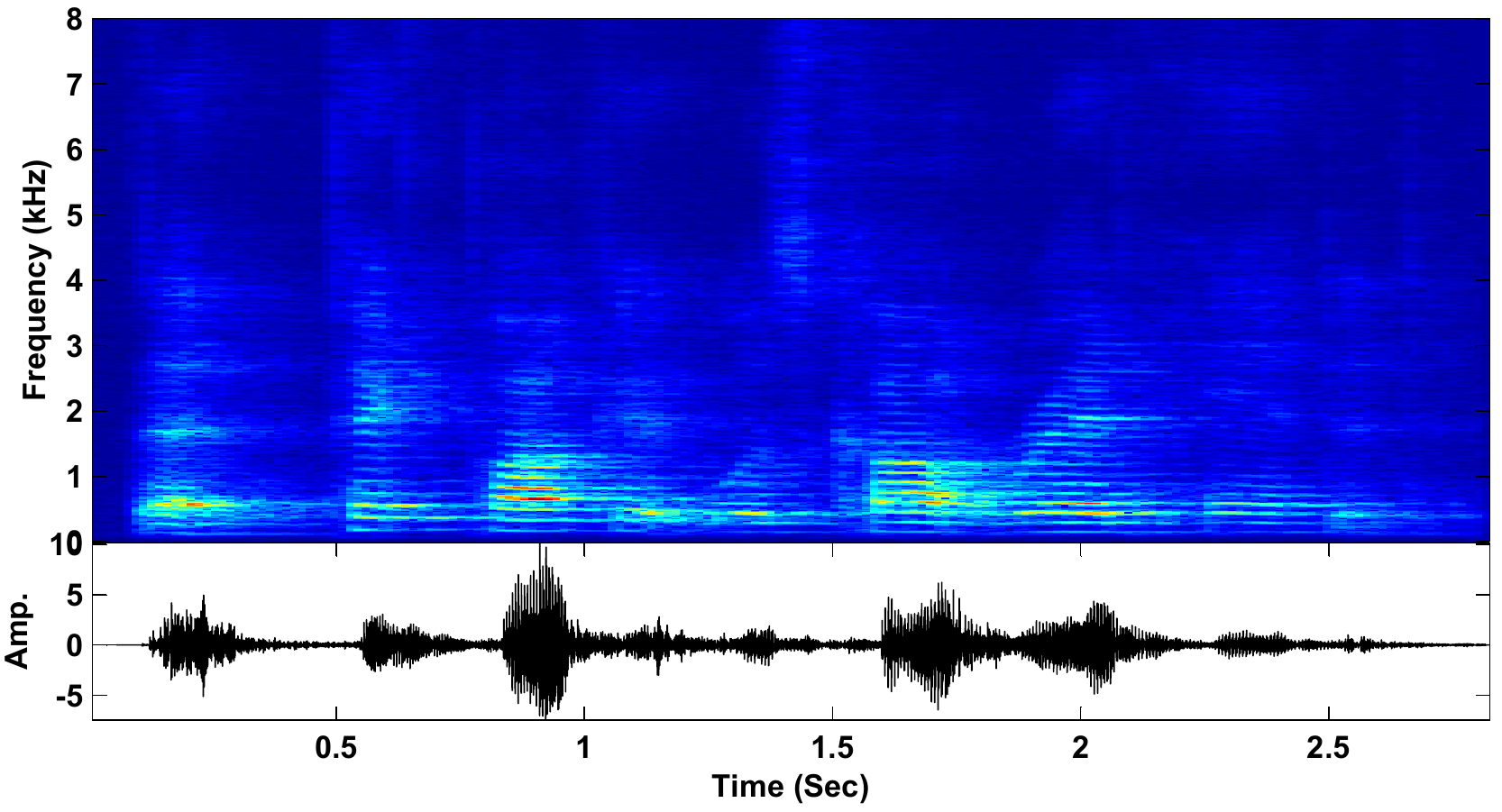}}

\subfloat[\label{fig:Dereverberated-speech-signal-using nctf-nmf-dicoc}Dereverberated
speech signal using N-CTF+NMF+OC]{\includegraphics[width=.995\columnwidth]{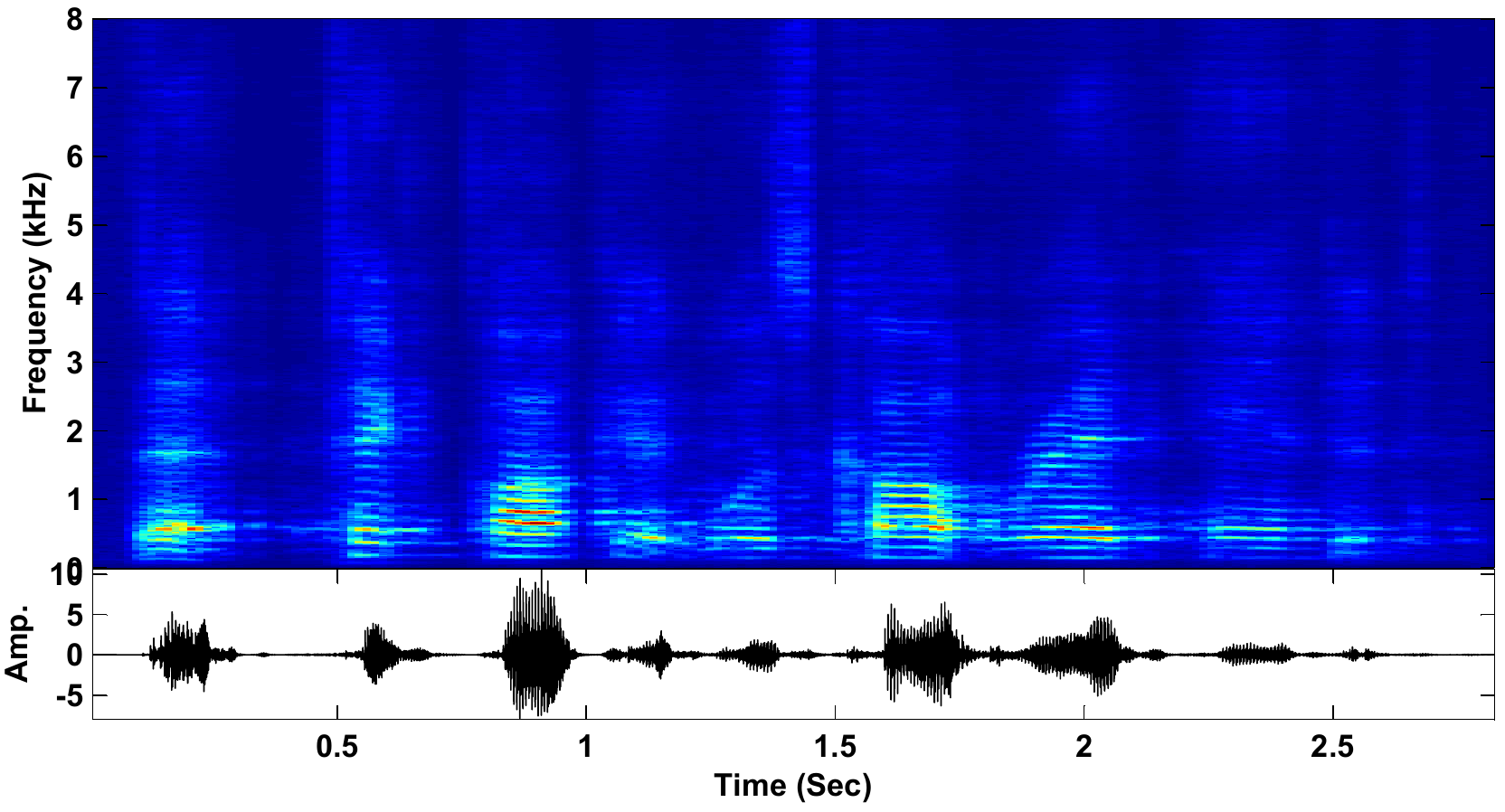}}

\subfloat[\label{fig:Dereverberated-speech-signal-using_se}Dereverberated speech
signal using SE ]{\includegraphics[width=.995\columnwidth]{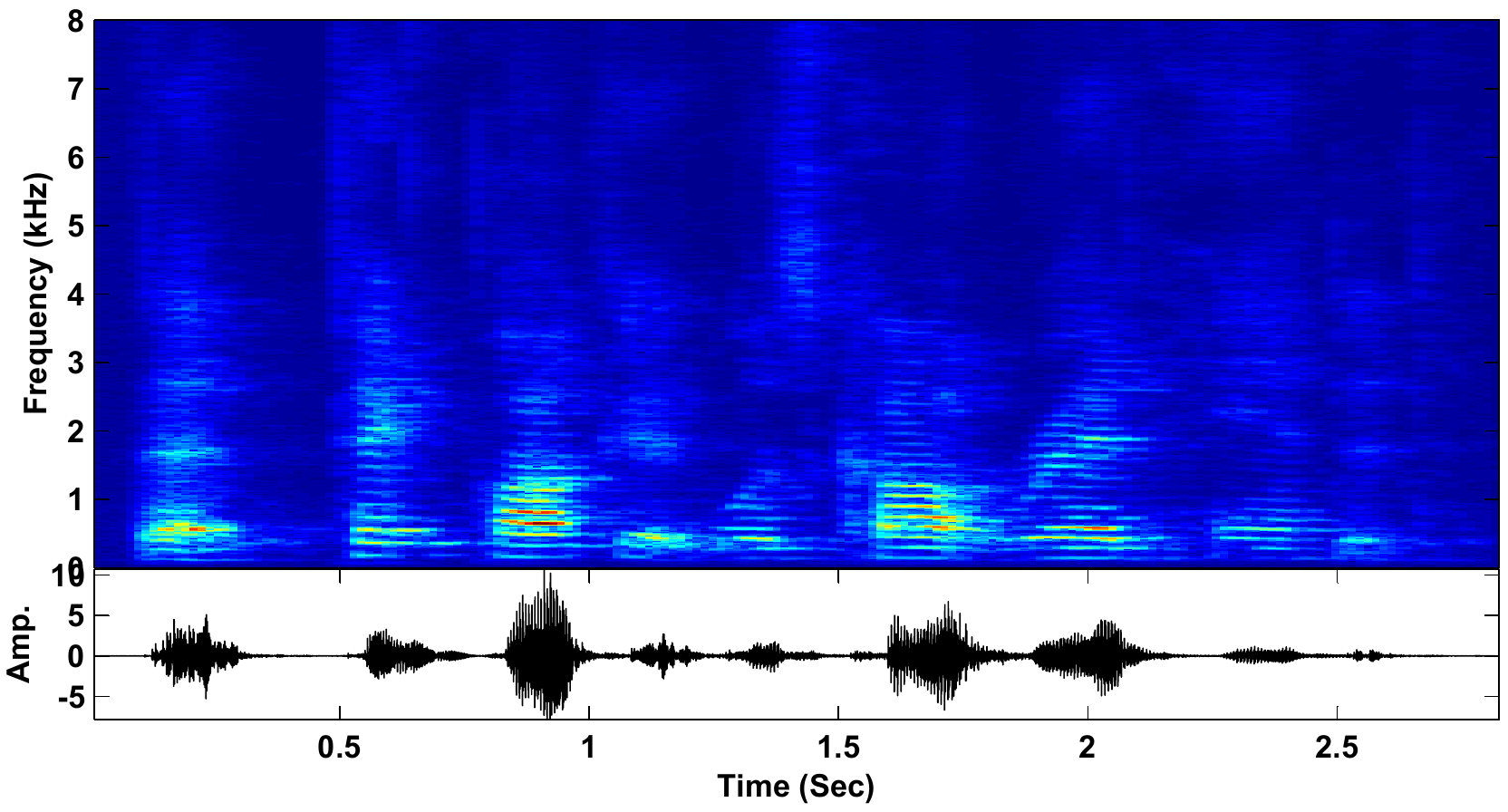}}

\subfloat[\label{fig:Clean-speech-signal}Clean speech signal]{\includegraphics[width=.995\columnwidth]{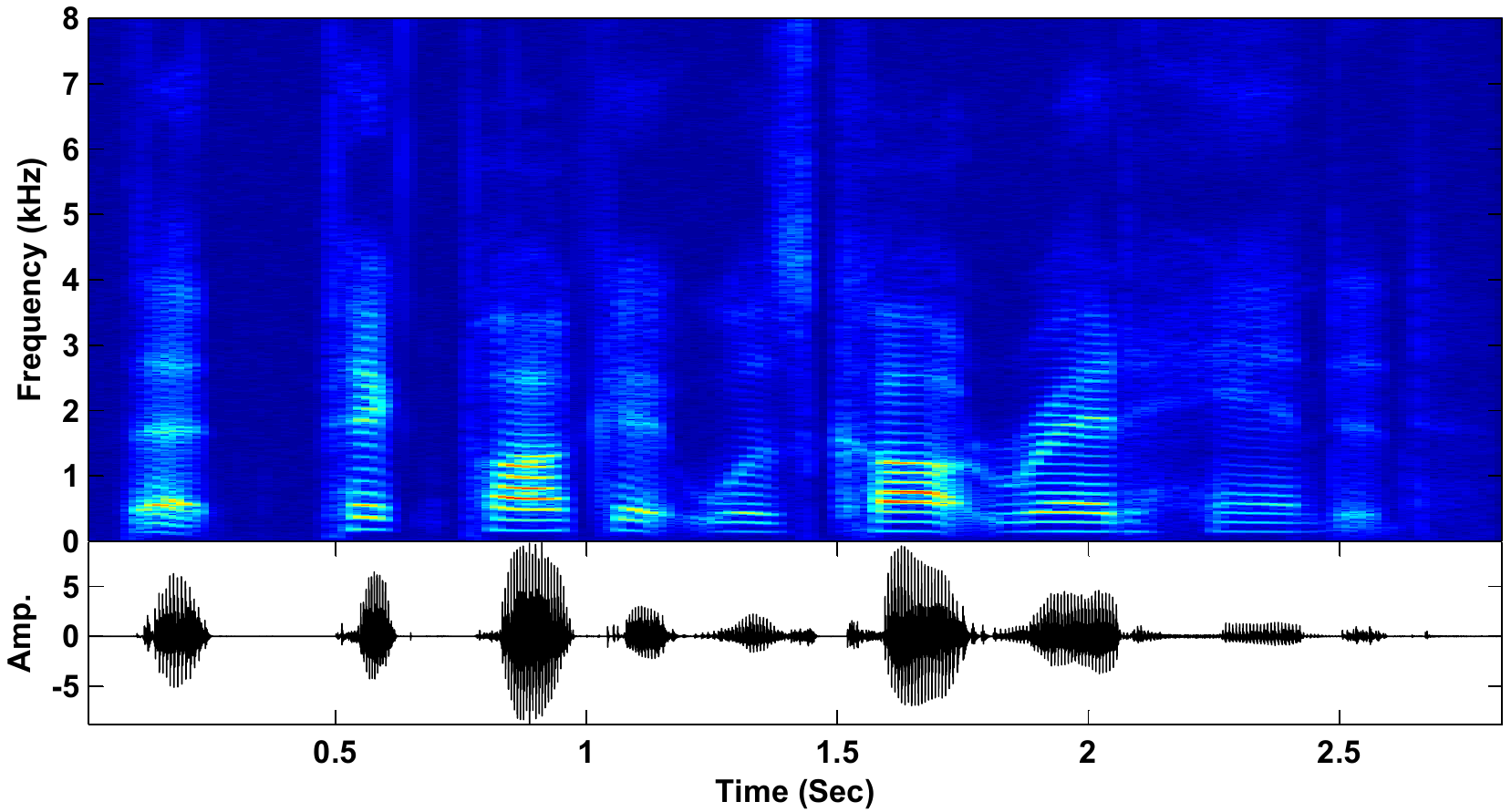}}

\caption{\label{fig:Spectrograms}Spectrograms of (a) reverberant microphone signal, (b) dereverberated signal using N-CTF+NMF+OC, (c) dereverberated signal using spectral
enhancement (SE), (d) clean speech signal ($T_{60}\approx680$ ms and $\text{DRR}\approx0$ dB).}
\end{figure}

%This section presents the experimental results using the integrated method.
Fig. \ref{fig:Spectrograms} depicts exemplary spectrograms (for the RIR with $T_{60}\approx680$ ms and $\text{DRR}\approx0$ dB) of the noiseless reverberant and clean speech signals together with the spectrograms
of the dereverberated speech signals using the proposed N-CTF+NMF+OC method and the SE method. As
can be observed, reverberation effects have been substantially reduced using the proposed method.
%Sound samples for this experiment are available at \url{https://dl.dropboxusercontent.com/u/80154985/nctf_nmf_audio.rar}.

To quantitatively compare the dereverberation performance in the absence of background noise, $\Delta\text{PESQ}$,
$\Delta\text{CD}$, and $\Delta\text{RDT}$ values obtained using several variants
of the integrated method, averaged over the 16 speech utterances, are shown in Fig. \ref{fig:PESQ-rir2},
Fig. \ref{fig:PESQ-rir3}, and Fig. \ref{fig:PESQ-rir4}  for the three
considered RIRs. Also, Fig.~\ref{fig:PESQ-rir_av} shows the average results over all 48 test utterances (3 RIRs and 16 speech utterances per RIR). In all figures, the suffix '(t)' is used to indicate that the methods use the
temporal dependencies. Several conclusions can be drawn by studying these
results:

%%%%%%%%%%%%%%%%%%%%%%%%%%%%%%%%%%%%%%%%%%%%%%%%%%%%%%%%%%%%%%%%%%%%%%%%%%%%%
\subsubsection{Using NMF-based spectral model in N-CTF}
%%%%%%%%%%%%%%%%%%%%%%%%%%%%%%%%%%%%%%%%%%%%%%%%%%%%%%%%%%%%%%%%%%%%%%%%%%%%%
By additionally using an NMF-based spectral model, i.e, using the
N-CTF+NMF method, the performance of the N-CTF-based dereverberation
method substantially improves for all RIRs. A consistent improvement
is observed for all three measures for all considered RIRs using
the N-CTF+NMF method compared to the N-CTF method.

%%%%%%%%%%%%%%%%%%%%%%%%%%%%%%%%%%%%%%%%%%%%%%%%%%%%%%%%%%%%%%%%%%%%%%%%%%%%%
\subsubsection{N-CTF-based methods versus spectral enhancement method}
%%%%%%%%%%%%%%%%%%%%%%%%%%%%%%%%%%%%%%%%%%%%%%%%%%%%%%%%%%%%%%%%%%%%%%%%%%%%%
The dereverberation method using only the N-CTF model leads to a higher
$\Delta\text{CD}$ but a lower $\Delta\text{RDT}$ compared to the spectral
enhancement (SE) method, which implies that the N-CTF method introduces
less distortion but also leaves more late reverberation in the dereverberated
speech signals. Considering PESQ, which evaluates the overall speech
quality, the N-CTF method results in slightly better scores compared
to the SE method. The proposed N-CTF+NMF method outperforms the SE
method for all considered RIRs and instrumental measures. For example, the proposed N-CTF+NMF method outperforms the SE method by more than 0.25 PESQ-MOS points for all considered RIRs.
\begin{figure}%[H]
\includegraphics[width=.995\columnwidth]{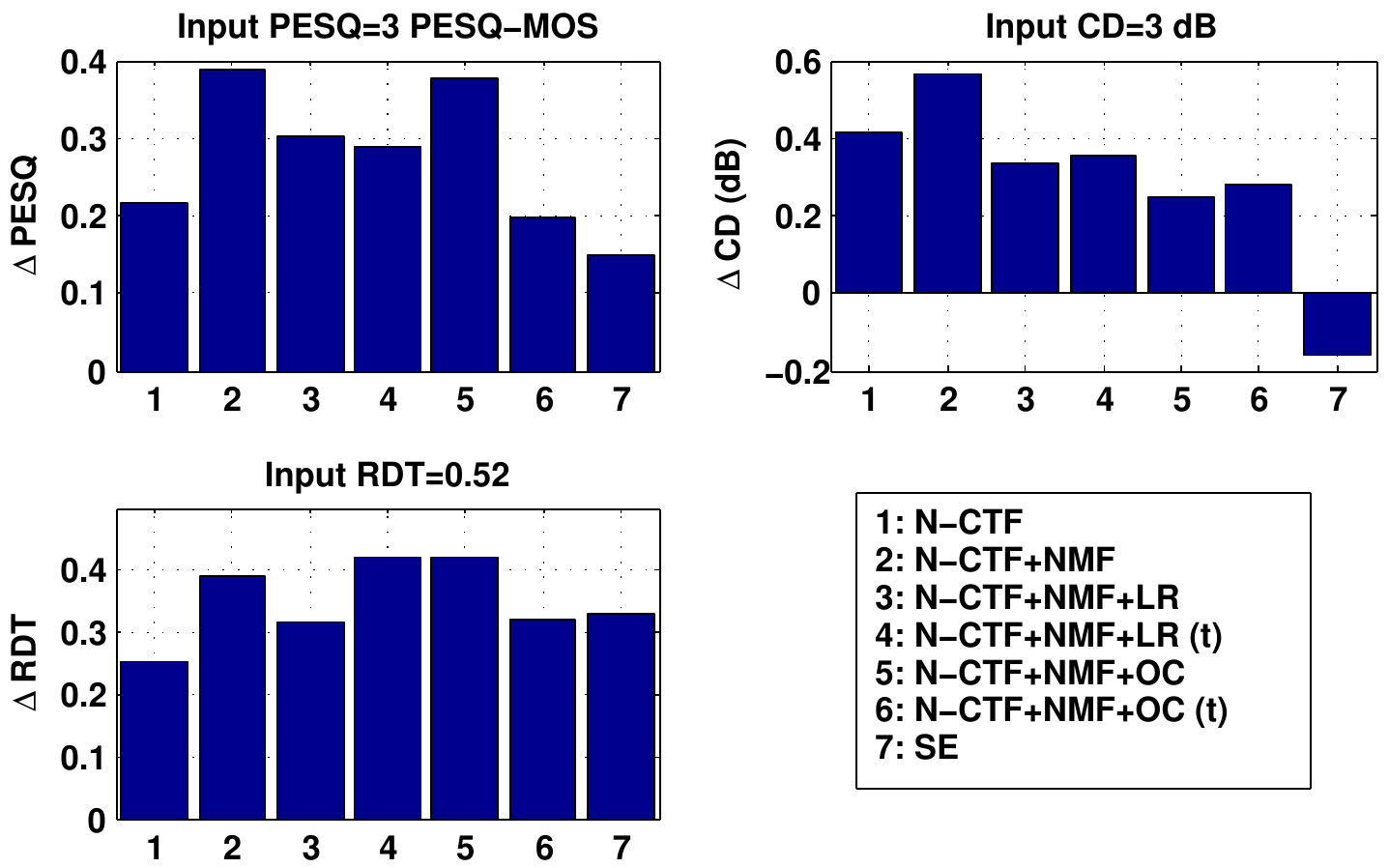}\caption{\label{fig:PESQ-rir2}Instrumental measures for the proposed methods
for a RIR with $T_{60}\approx640$ ms and $\text{DRR}\approx12$ dB.}
\end{figure}

\begin{figure}%[H]
\includegraphics[width=.995\columnwidth]{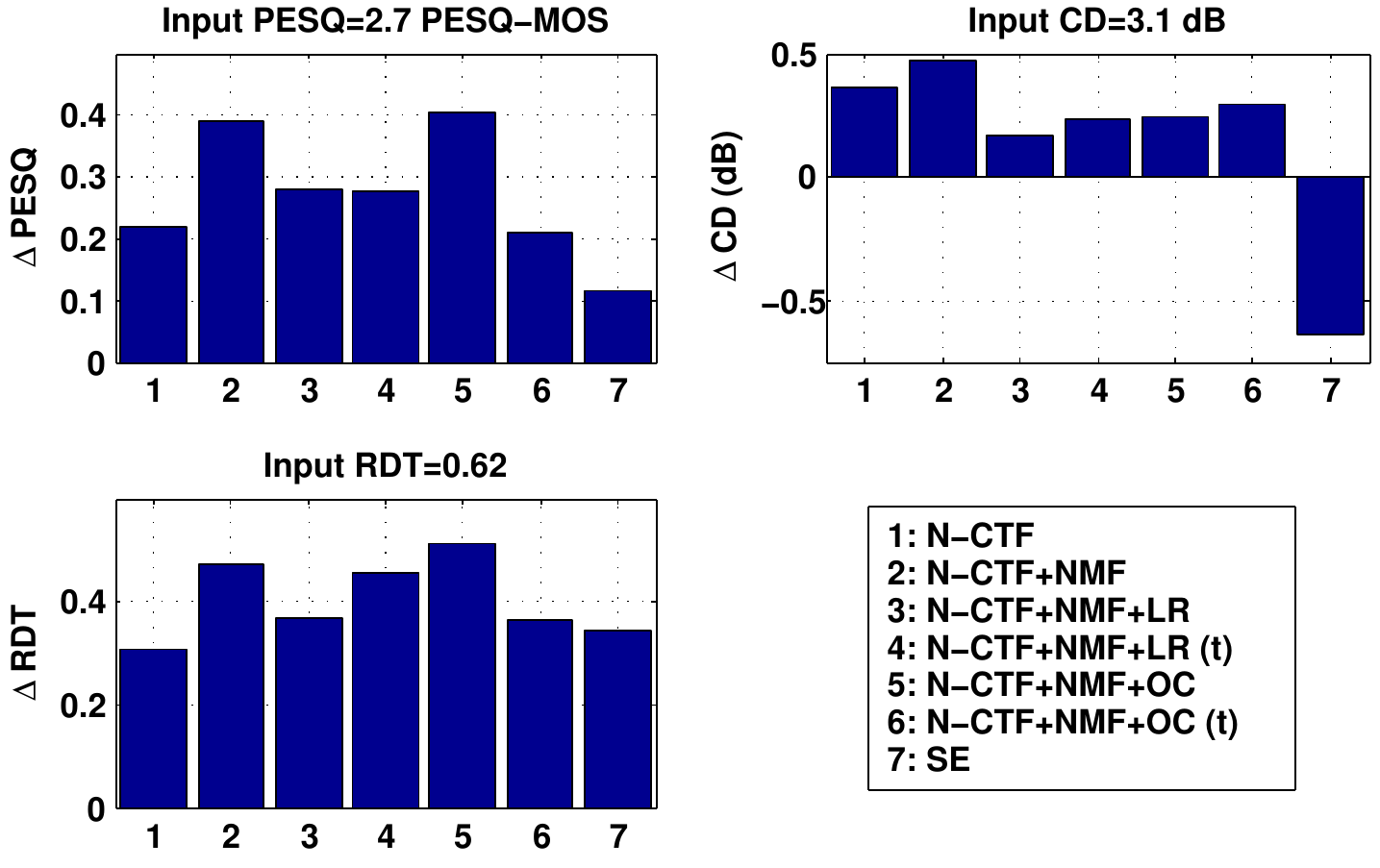}\caption{\label{fig:PESQ-rir3}Instrumental measures for the proposed methods
for a RIR with $T_{60}\approx430$ ms and $\text{DRR}\approx5$ dB.}
\end{figure}

\begin{figure}
\includegraphics[width=.995\columnwidth]{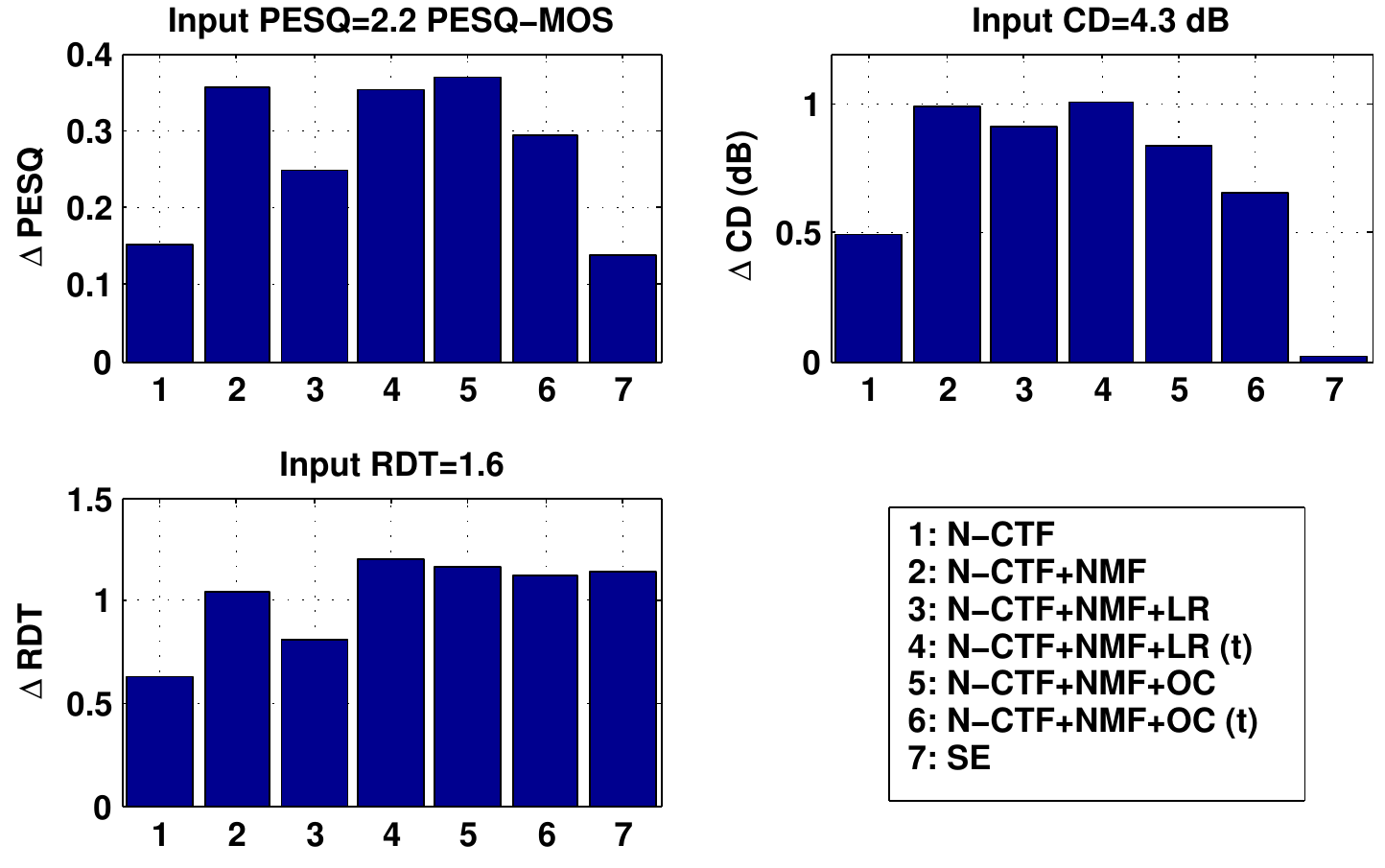}\caption{\label{fig:PESQ-rir4}Instrumental measures for the proposed methods
for a RIR with $T_{60}\approx680$ ms and $\text{DRR}\approx0$ dB.}
\end{figure}

\begin{figure}
\includegraphics[width=.995\columnwidth]{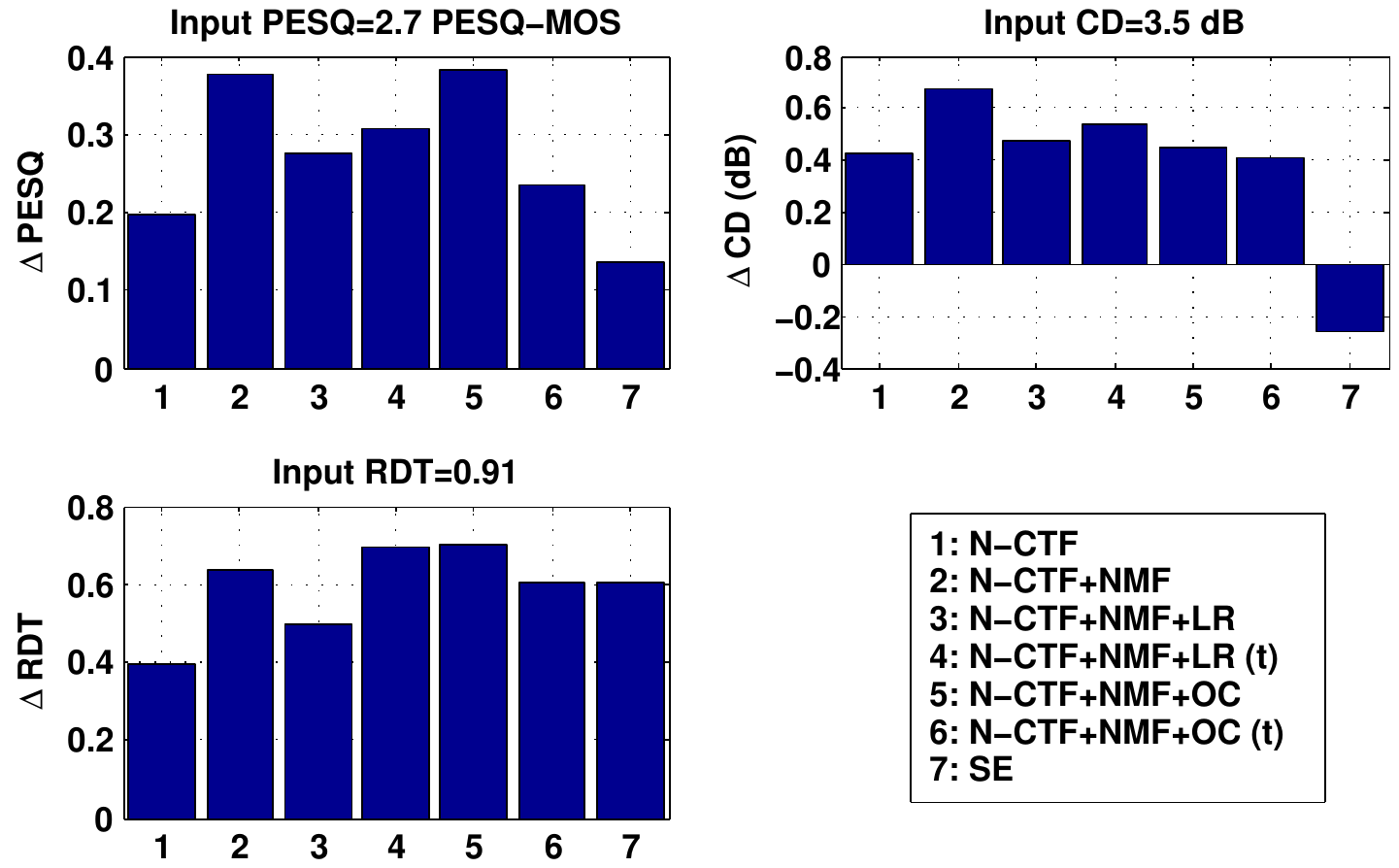}\caption{\label{fig:PESQ-rir_av}Instrumental measures for the proposed methods, averaged over the three considered RIRs.}
\end{figure}
%%%%%%%%%%%%%%%%%%%%%%%%%%%%%%%%%%%%%%%%%%%%%%%%%%%%%%%%%%%%%%%%%%%%%%%%%%%%%
\subsubsection{Offline- versus online-learned basis matrices}
%%%%%%%%%%%%%%%%%%%%%%%%%%%%%%%%%%%%%%%%%%%%%%%%%%%%%%%%%%%%%%%%%%%%%%%%%%%%%
The results consistently show that the performance of the N-CTF+NMF+LR\textit{
}method with $R=100$ offline-learned basis vectors is worse than
the performance of the online counterpart N-CTF+NMF. However, by using
a larger basis matrix with $R=3000$ offline-learned basis vectors (N-CTF+NMF+OC) the performance improves and
is comparable to the performance of the N-CTF+NMF method. Both these methods result in comparable $\Delta\text{PESQ}$ scores; the N-CTF+NMF+OC
method results in a lower $\Delta\text{CD}$ but a higher $\Delta\text{RDT}$
compared to the N-CTF+NMF method. This can be explained by the fact
that in the N-CTF+NMF+OC method the NMF basis matrix is learned offline using speaker-independent training data and
is held fixed. This may lead to some artifacts in the dereverberated
speech signal due to the mismatch between training and testing data.
At the same time, since the basis matrix is held fixed, the model
is able to better separate the reverberant component from the clean
component, which leads to a larger reduction of the reverberant tail.

%%%%%%%%%%%%%%%%%%%%%%%%%%%%%%%%%%%%%%%%%%%%%%%%%%%%%%%%%%%%%%%%%%%%%%%%%%%%%
\subsubsection{Using temporal dependencies}
%%%%%%%%%%%%%%%%%%%%%%%%%%%%%%%%%%%%%%%%%%%%%%%%%%%%%%%%%%%%%%%%%%%%%%%%%%%%%
The experiments show that modeling the temporal dependencies using the frame-stacking method only appears to be useful
for highly reverberant conditions when the N-CTF+NMF+LR method is used. % (see Fig. \ref{fig:PESQ-rir4}).
%To model the temporal dependencies, $T_{st}=6$ (in (\ref{eq:update_h_united_stacked}))
%consecutive frames were stacked, and the high-dimensional basis matrix
%$\mathbf{W}$ was learned offline from clean speech training data.
As can be seen in Fig. \ref{fig:PESQ-rir4},
using the temporal dependencies in the N-CTF+NMF+LR method (i.e.,
the N-CTF+NMF+LR (t) method) leads to an additional improvement in the
instrumental measures. By using the temporal dependencies the reverberant
tail is significantly reduced but the cepstral distance is not changed
noticeably. However, for the N-CTF+NMF+OC method, no improvement is observed
by additionally using the temporal dependencies (i.e., using the N-CTF+NMF+OC
(t) method).
%Note that modeling temporal dependencies is better justified when using offline-learned basis matrices than online-learned basis matrices because learning .
It should be mentioned that the N-CTF+NMF (t) method (not shown in the figures) resulted in a similar performance as the N-CTF+NMF method, and hence it is omitted in the figures for readability.

\begin{figure}
\includegraphics[width=.995\columnwidth]{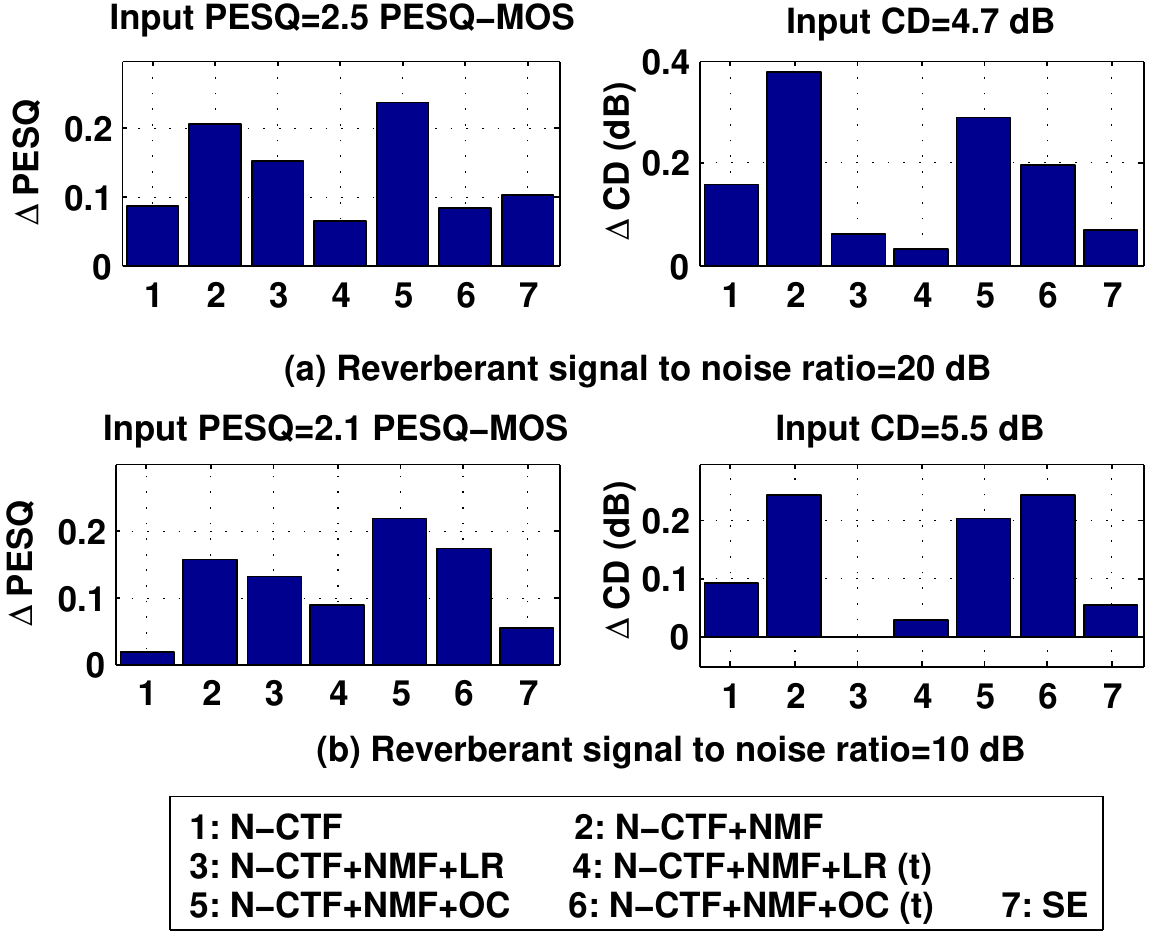}\caption{\label{fig:rir_noise_av}Instrumental measures for the proposed methods for two background noise levels, averaged over the three considered RIRs.}
\end{figure}

The experimental results shown in Fig.~\ref{fig:PESQ-rir2}-\ref{fig:PESQ-rir_av} are obtained without considering any background noise. To investigate the robustness of the proposed dereverberation methods to the presence of background noise, experiments were performed where speech-shaped noise was added to the reverberant microphone signals at two reverberant-signal-to-noise ratios (10 dB and 20 dB). Fig.~\ref{fig:rir_noise_av} shows the average $\Delta\text{PESQ}$ and $\Delta\text{CD}$ values over all 48 test utterances. Due to space limitation, results for individual RIRs are not shown in the paper. It should be noted that due to the presence of background noise the input $\text{PESQ}$ and $\text{CD}$ scores are lower in this figure compared to Fig.~\ref{fig:PESQ-rir_av}. The results show that the dereverberation performance of all the  methods degrade when a background noise is added to the reverberant microphone signals.  As can be seen in Fig.~\ref{fig:rir_noise_av}, the best performance is obtained using the N-CTF+NMF and N-CTF+NMF+OC methods. These experiments show that the proposed methods are quite robust to the presence of background noise, and that even in the noisy scenarios, the proposed methods result in higher scores for the dereverberated speech signals compared to the baseline N-CTF and SE methods. It is important to note that we have not explicitly modeled the background noise in the proposed methods. This can be an interesting extension of the current methods, where Eq.~\eqref{eq:N-CTF} is modified such that the background noise is also explicitly modeled, e.g., using a separate NMF model.

%\begin{figure}
%\includegraphics[width=.995\columnwidth]{figures/rir3_noise}\caption{\label{fig:rir3_noise}$\Delta\text{PESQ}$ and $\Delta\text{CD}$ for the proposed methods
%for a RIR with $T_{60}\approx680$ ms and $\text{DRR}\approx0$ dB, for two background noise levels.}
%\end{figure}
%}

%%%%%%%%%%%%%%%%%%%%%%%%%%%%%%%%%%%%%%%%%%%%%%%%%%%%%%%%%%%%%%%%%%%%%%%%%%%%%
\subsection{Weighted Method}
\begin{figure}
\includegraphics[width=.995\columnwidth]{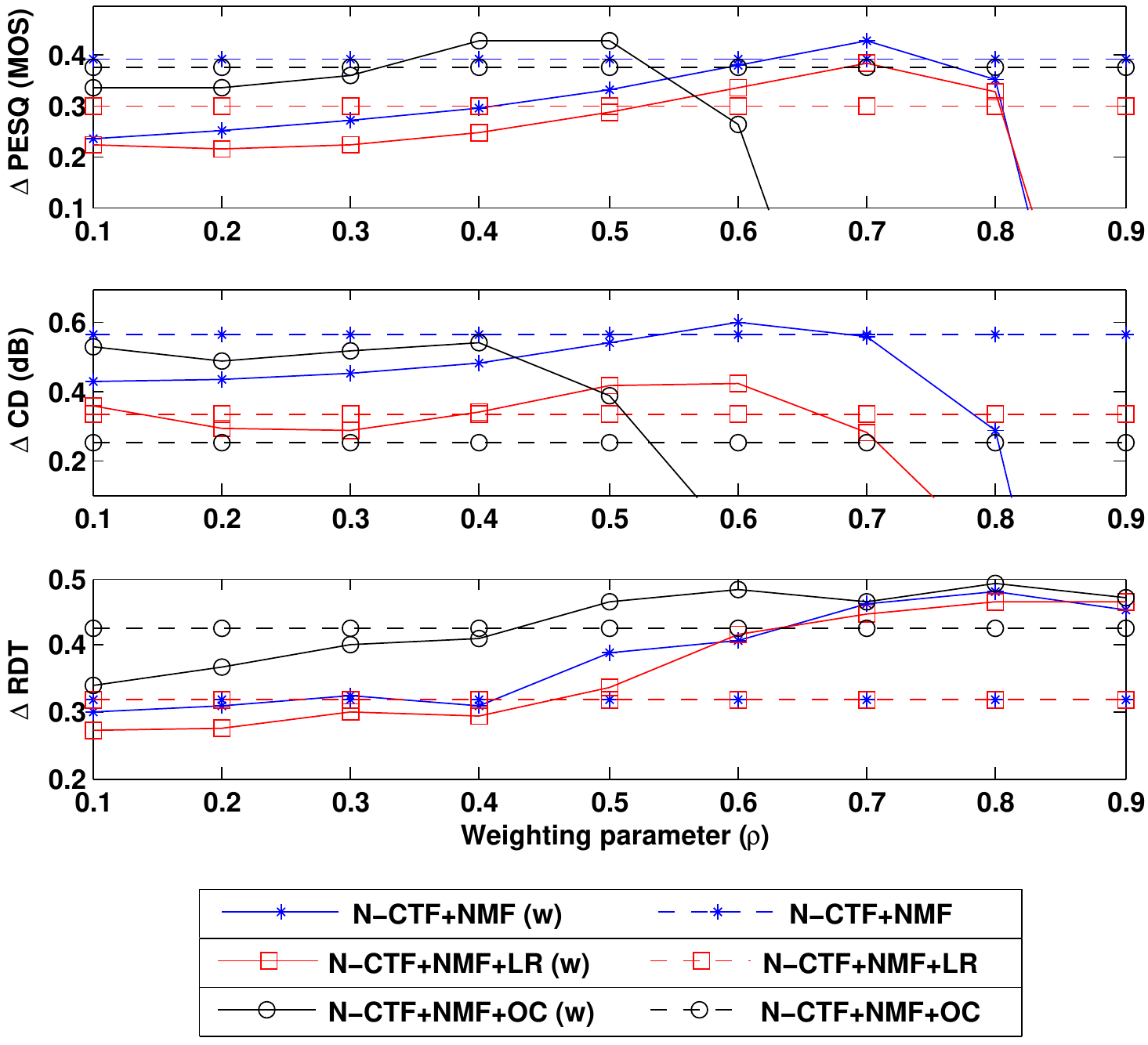}\caption{\label{fig:PESQ-weighted-rir2}Instrumental measures for the proposed
integrated and weighted methods for a RIR with $T_{60}\approx640$
ms and $\text{DRR}\approx12$ dB.}
\end{figure}

\begin{figure}
\includegraphics[width=.995\columnwidth]{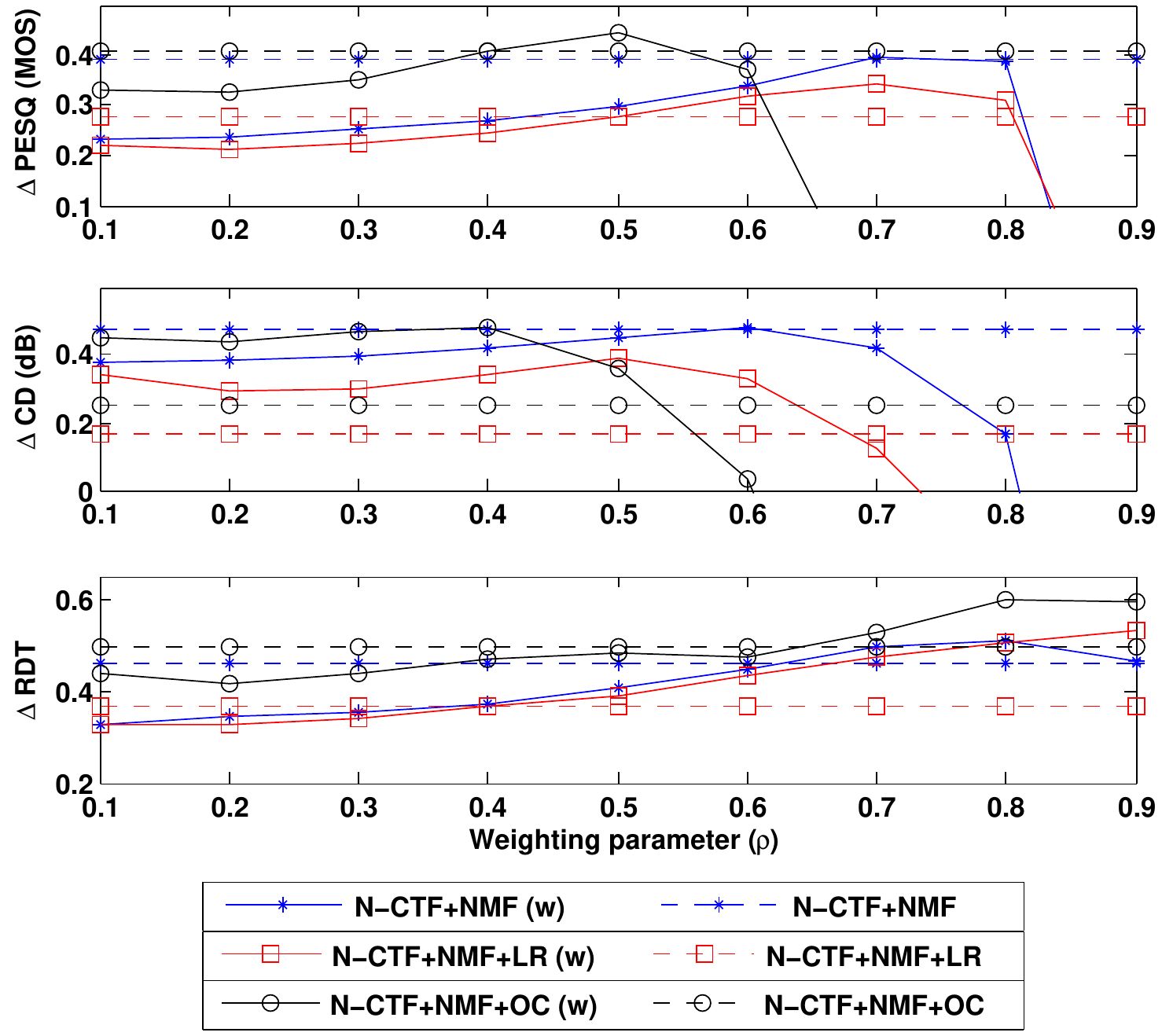}\caption{\label{fig:PESQ-weighted-rir3}Instrumental measures for the proposed
integrated and weighted methods for a RIR with $T_{60}\approx430$
ms and $\text{DRR}\approx5$ dB.}
\end{figure}

\begin{figure}
\includegraphics[width=.995\columnwidth]{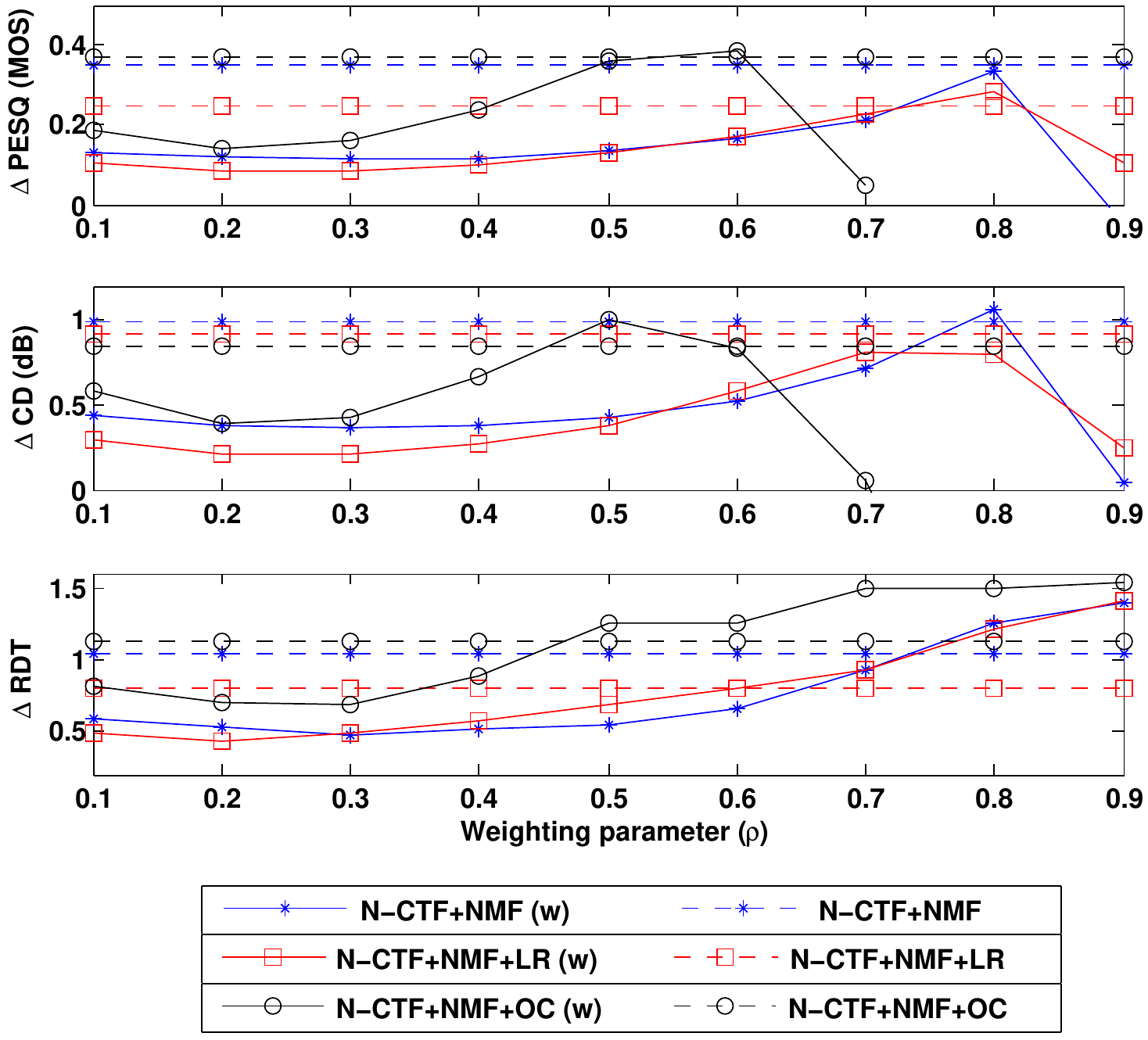}\caption{\label{fig:PESQ-weighted-rir4}Instrumental measures for the proposed
integrated and weighted methods for a RIR with $T_{60}\approx680$
ms and $\text{DRR}\approx0$ dB.}
\end{figure}

%%%%%%%%%%%%%%%%%%%%%%%%%%%%%%%%%%%%%%%%%%%%%%%%%%%%%%%%%%%%%%%%%%%%%%%%%%%%%
In this section the performance of the proposed integrated
and weighted methods is compared, where no background noise is present. Fig. \ref{fig:PESQ-weighted-rir2},
Fig. \ref{fig:PESQ-weighted-rir3}, and Fig. \ref{fig:PESQ-weighted-rir4}
show the $\Delta\text{PESQ}$, $\Delta\text{CD}$, and $\Delta\text{RDT}$
obtained using the integrated and weighted methods
as a function of the weighting parameter $\rho$ for different RIRs. These figures compare
the performance of three variants of the integrated method
(N-CTF+NMF, N-CTF+NMF+LR, and N-CTF+NMF+OC) to the performance of
the same variants of the weighted method, which are identified
by the suffix '(w)'.

Results show that an optimal value for the weighting parameter $\rho$
(denoted by $\rho^{\star}$) can be found, using which the obtained $\Delta\text{PESQ}$
for the weighted methods is similar or slightly better
than the obtained $\Delta\text{PESQ}$ for the integrated methods. However, the optimal weighting parameter $\rho^{\star}$ depends on the considered RIR and the variant of the weighted method, i.e. the utilized NMF model. The results also show that the $\Delta\text{CD}$ measure is substantially
higher for the N-CTF+NMF+OC (w) method using the optimal $\rho^{\star}$
compared to the N-CTF+NMF+OC method, while the two methods result
in similar $\Delta\text{RDT}$ values. Moreover, using the optimal
$\rho^{\star}$, the obtained $\Delta\text{CD}$ and $\Delta\text{RDT}$
values using the N-CTF+NMF (w) and N-CTF+NMF+LR (w) methods are similar
to the $\Delta\text{CD}$ and $\Delta\text{RDT}$ values obtained
using the N-CTF+NMF and N-CTF+NMF+LR methods.

Although it is possible to achieve a better dereverberation performance using the weighted method, the experiments show that the optimal weighting parameter $\rho^{\star}$
highly depends on both the room acoustics as well as the NMF spectral model, which is a disadvantage compared to the integrated method. It can be seen that for the N-CTF+NMF+OC (w) method, $\rho^{\star}$ is around $0.4-0.5$,
while for the N-CTF+NMF (w) and N-CTF+NMF+LR (w) methods, $\rho^{\star}$ is around $0.7-0.8$.
By increasing the value of $\rho$, a lower reverberant tail, i.e., higher
$\Delta\text{RDT}$, and a larger spectral distance, i.e., lower $\Delta\text{CD}$, are typically achieved. This is more notable when an overcomplete NMF basis matrix
$\mathbf{W}$ is learned (N-CTF+NMF+OC (w)).
This may be explained by noting that, in this case, the columns of
$\mathbf{W}$ are sampled from the spectrograms of the speaker-independent
training clean speech signals. Therefore, each spectral vector of
the dereverberated speech signal is approximated using very few, strictly
speaking only one, columns of $\mathbf{W}$. This leads to a large
value for the NMF cost function $P$ in (\ref{eq:combined}), and
accordingly a relatively small $\rho$ leads to the best dereverberation
performance. For larger values of $\rho$ (especially when $\rho>0.8$) an estimate of the clean speech spectrogram is obtained that is a combination
of the clean spectral vectors, and hence, a higher $\Delta\text{RDT}$ value
is obtained. At the same time, since a lower weight is
given to the acoustic cost function $Q$ in (\ref{eq:combined}),
the obtained estimate of the clean speech spectrogram is highly distorted
because it is largely independent of the observed signal.

%%%%%%%%%%%%%%%%%%%%%%%%%%%%%%%%%%%%%%%%%%%%%%%%%%%%%%%%%%%%%%%%%%%%%%%%%%%%%
\subsection{Influence of the Parameters\label{sub:Important-Parameters}}
%%%%%%%%%%%%%%%%%%%%%%%%%%%%%%%%%%%%%%%%%%%%%%%%%%%%%%%%%%%%%%%%%%%%%%%%%%%%%
In this section the influence of different parameters on the speech dereverberation performance in the absence of background noise is investigated. First, experiments show that the parameter $p$, which
determines whether magnitude or power spectrograms are used, has a significant
effect on the performance. Additionally,
the number of iterations for the update rules and the
STFT frame length were found to be quite influential. The effect of
these parameters on the performance of the integrated method
is separately studied, where only one parameter is
varied and the other parameters are set to the values mentioned in
Section \ref{sub:Evaluation_and_Implementation}, i.e., $p=1$ (magnitude
spectrogram), frame length = 64 ms, and number of iterations = 20.
The experiments in this section are performed for the RIR with $T_{60}\approx680$
ms and $\text{DRR}\approx0$ dB, but similar observations were made for the other RIRs.

\begin{figure}
\includegraphics[width=.995\columnwidth]{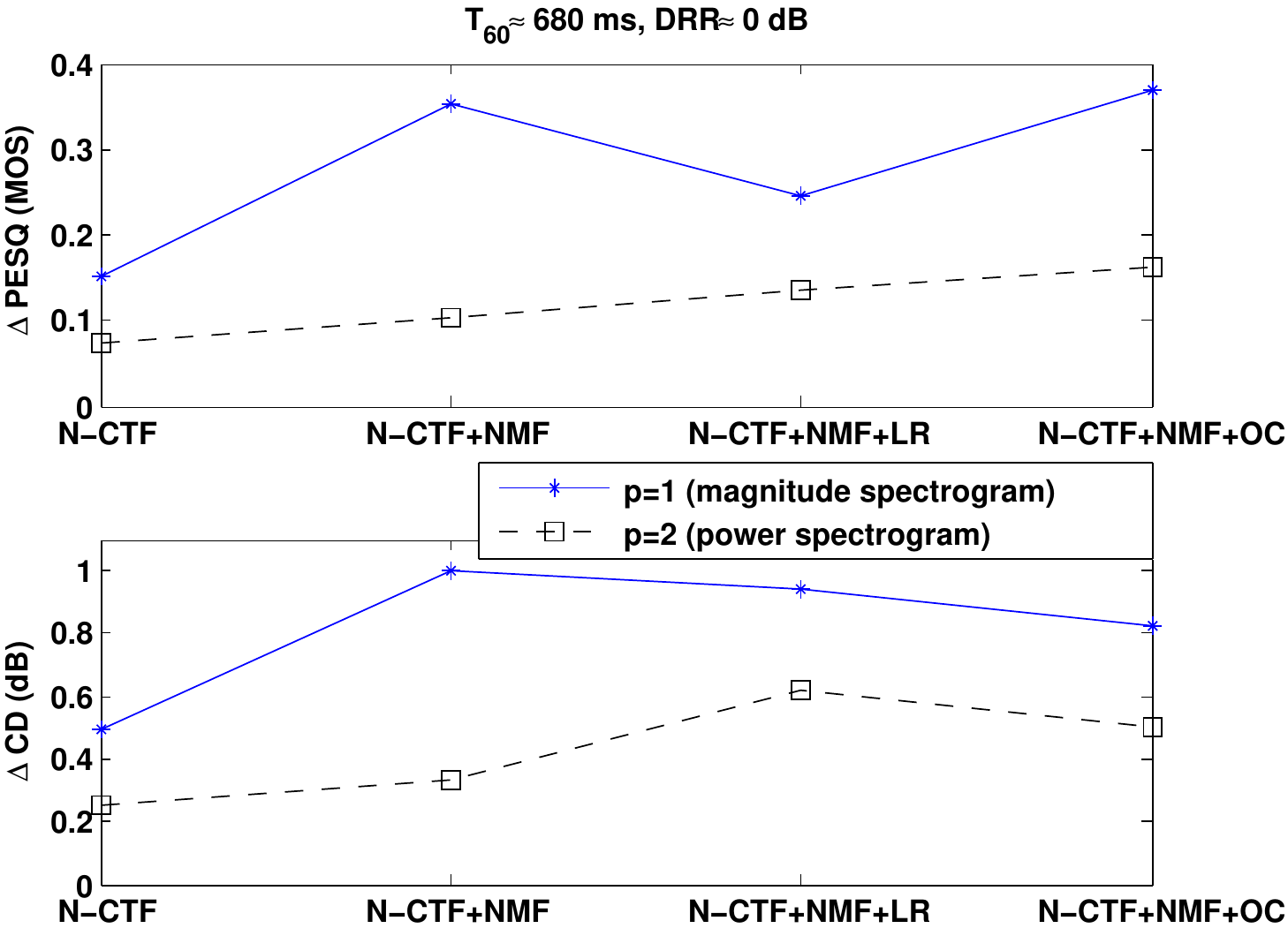}\caption{\label{fig:sim_power}Dereverberation performance using magnitude spectrogram ($p=1$)
and power spectrogram ($p=2$).}
\end{figure}

\textbf{1) Magnitude versus power spectrogram. }Fig. \ref{fig:sim_power}
shows $\Delta\text{PESQ}$ and $\Delta\text{CD}$ for power and magnitude
spectrograms using the integrated methods. As can be clearly seen, a better
performance is obtained using the magnitude spectrogram for all variants, even though the N-CTF model in (\ref{eq:N-CTF}) can theoretically
be better justified using the power spectrogram. However, a similar observation has
already been also made in NMF-based source separation and speech enhancement
\cite{Mohammadiha2013g}.

\begin{figure}
\includegraphics[width=.995\columnwidth]{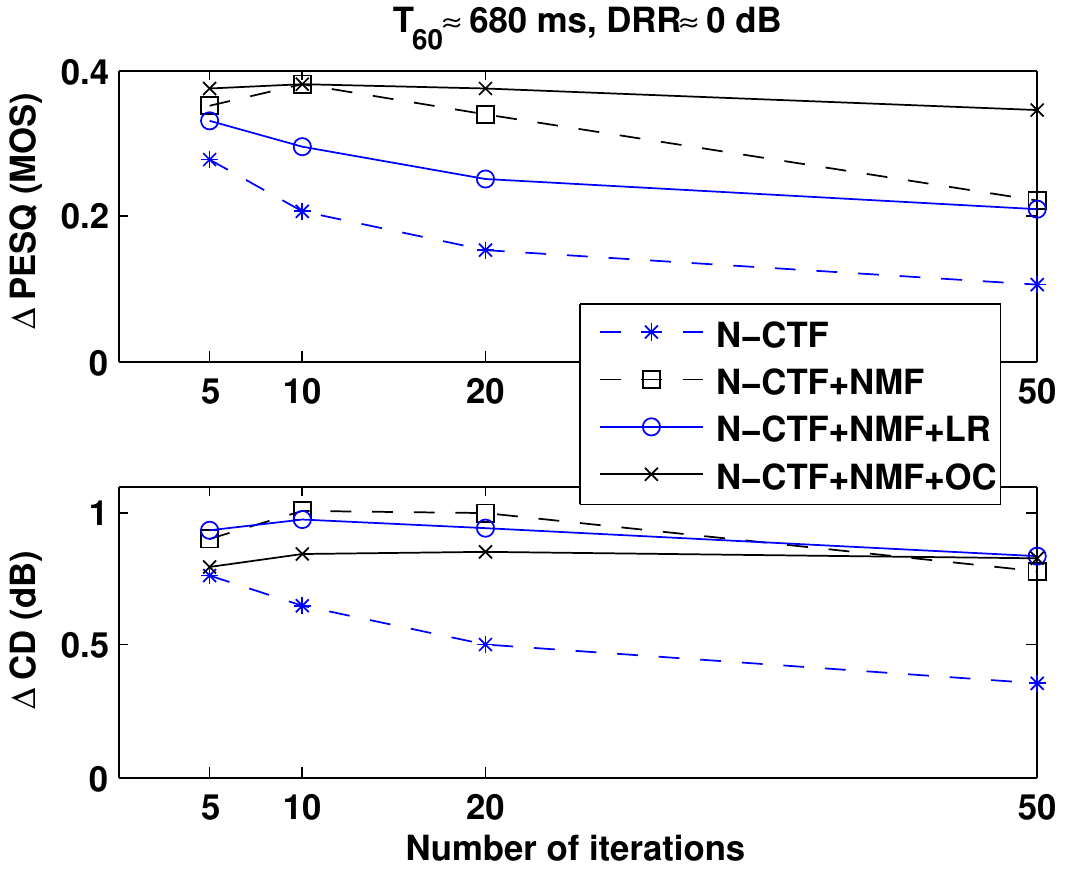}\caption{\label{fig:sim_num_iter}Dereverberation performance as a function
of the number of iterations.}
\end{figure}

\textbf{2) Number of iterations}. Fig. \ref{fig:sim_num_iter} shows $\Delta\text{PESQ}$ and $\Delta\text{CD}$ as a function of the
number of iterations, where the $\text{number of iterations}\in\{5,10,20,50\}$.
Results show that a small number of iterations (in the range of 5-20)
is enough to obtain the best dereverberation performance.

\begin{figure}
\includegraphics[width=.995\columnwidth]{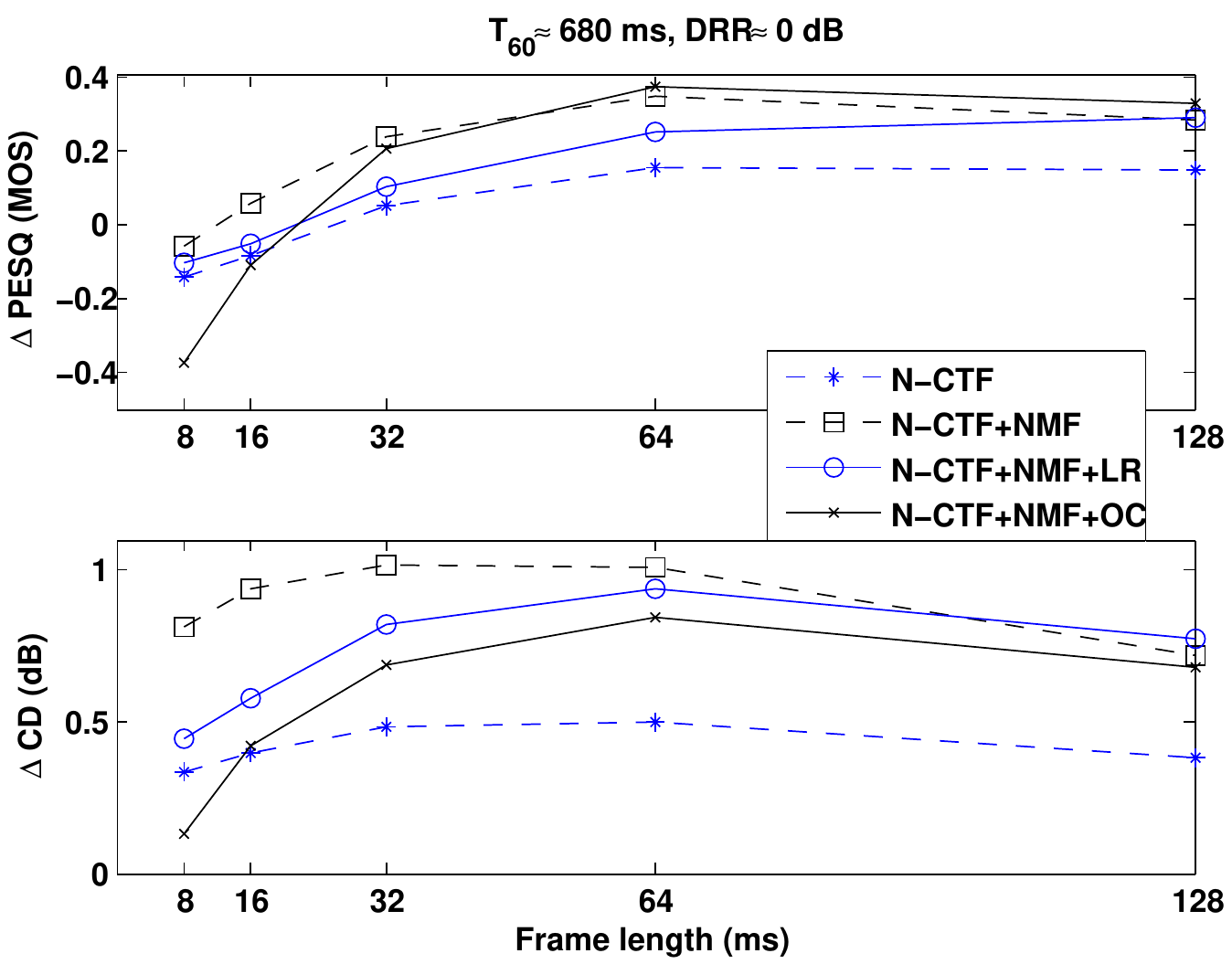}\caption{\label{fig:sim_frameLength}Dereverberation performance as a function
of the STFT frame length using 50\% overlapping square-root Hann windows. }
\end{figure}

\textbf{3) Frame length. }Fig. \ref{fig:sim_frameLength} shows $\Delta\text{PESQ}$
and $\Delta\text{CD}$ as a function of the STFT frame length, where
the $\text{frame length}\in\{8,16,32,64,128\}$ ms and 50\%-overlapping
square-root Hann windows are used for all cases. As can be seen, the
performance degrades significantly when shorter frames are used. The
best performance is obtained when the frame length is around 64 ms.

%%%%%%%%%%%%%%%%%%%%%%%%%%%%%%%%%%%%%%%%%%%%%%%%%%%%%%%%%%%%%%%%%%%%%%%%%%%%%
\section{Conclusion\label{sec:Conclusion}}
%%%%%%%%%%%%%%%%%%%%%%%%%%%%%%%%%%%%%%%%%%%%%%%%%%%%%%%%%%%%%%%%%%%%%%%%%%%%%
In this paper, we have considered single-channel speech dereverberation methods combining an N-CTF-based acoustic model and an NMF-based spectral model
in order to jointly exploit the room acoustics and speech spectral structure.
Two methods are presented to combine the N-CTF and NMF models, namely the integrated method, where the NMF model is directly integrated into the N-CTF model, and the weighted
method, where the N-CTF and NMF cost functions are weighted and summed. For both methods, generalized Kullback-Leibler divergence is used to define the cost functions and iterative update rules are derived to estimate the clean speech spectrogram.

Experimental results, with and without background noise, for three different RIRs showed that the performance of the N-CTF-based
dereverberation method % (measured using PESQ, cepstral distance, and reverberation decay tail)
was significantly improved by additionally exploiting the NMF-based spectral model, where considerably better-quality speech signals were obtained
using the magnitude spectrograms compared to the power spectrograms. It was shown that the integrated method outperforms a state-of-the-art spectral
enhancement method by 0.25 PESQ-MOS points.
Results also showed that using the weighted method it is possible to achieve even a better performance, but that the optimal weighting parameter highly depends on the NMF model as well as the room acoustics.
Using temporal dependencies based on a frame-stacking method
was found to be  useful only for highly reverberant conditions when a
low-rank NMF basis matrix was learned offline from clean speech training
data.
%Future work will focus on also modeling the background noise in the proposed dereverberation methods.\nminsert{effect of background noise}
%%%%%%%%%%%%%%%%%%%%%%%%%%%%%%%%%%%%%%%%%%%%%%%%%%%%%%%%%%%%%%%%%%%%%%%%%%
\appendices
%%%%%%%%%%%%%%%%%%%%%%%%%%%%%%%%%%%%%%%%%%%%%%%%%%%%%%%%%%%%%%%%%%%%%%%%%%
\section{Assumptions Underlying the N-CTF Model\label{app: Appendix_NCTF_assumptions}}
Assuming that the phases of $h_{c}\left(k,\tau\right)$ at different frames are mutually independent uniformly-distributed random variables, (\ref{eq:CTF-complex}) leads to \cite{Kameoka2009}:
\begin{equation}
E\left(\left|y_{c}\left(k,t\right)\right|^{2}\right)\approx\sum_{\tau=0}^{L_{h}-1}\left|h_{c}\left(k,\tau\right)\right|^{2}\left|s_{c}\left(k,t-\tau\right)\right|^{2},\label{eq:expected_ctf}
\end{equation}
where $E\left(\cdot\right)$ denotes the mathematical expectation operator,
and $\left|\cdot\right|$ denotes absolute value operator. Although this assumption about the phase may seem unrealistic, it
is interesting to note that a similar expression has also been used in other state-of-the-art methods, such as
\cite{Habets2009}, to relate the spectral variance of the reverberant
speech signal to the spectral variance of the clean speech signal. In \cite{Habets2009}, assuming an exponential-decay model for the complex-valued RIR spectral coefficients $h_c(k,\tau)$, and assuming that $h_c(k,\tau)$ at different frames are mutually independent and zero-mean random variables with Gaussian distributions, it is shown that
\begin{equation}
E\left(\left|y_{c}\left(k,t\right)\right|^{2}\right)\approx\sum_{\tau=0}^{\infty}E\left(\left|h_{c}\left(k,\tau\right)\right|^{2}\right)E\left(\left|s_{c}\left(k,t-\tau\right)\right|^{2}\right),\label{eq:habets_model}
\end{equation}
where it is additionally assumed that the speech spectral coefficients $s_c(k,t)$ are independent and identically distributed zero-mean complex random variables with a certain distribution, and that the speech and the RIR spectral coefficients $s_c(k,t)$ and $h_c(k,\tau)$ are mutually independent.

Expression \eqref{eq:expected_ctf} is similar to \eqref{eq:habets_model} in that they both describe the spectral variance of the reverberant
speech signal as a convolution of two non-negative signals, which are related to the magnitude-squared RIR and speech spectral coefficients $|h_c(k,\tau)|^{2}$ and $|s_c(k,t)|^{2}$. These expression differ in that instantaneous magnitude-squared coefficients are used in \eqref{eq:expected_ctf} while expected magnitude-squared coefficients are used in \eqref{eq:habets_model}.
%%%%%%%%%%%%%%%%%%%%%%%%%%%%%%%%%%%%%%%%%%%%%%%%%%%%%%%%%%%%%%%%%%%%%%%%%%%%%%%%
\section{Iterative Learning Using Auxiliary Functions\label{app:iterative_learning}}
Iterative learning is a commonly used method to minimize a cost function with non-negativity constraints. This section provides a brief review of an iterative learning method based on the auxiliary function method \cite{Lee2000} that is used in this paper.

%\textbf{Iterative Learning:}
Consider a cost function $Q(h,s)$, where the unknown
parameters $h$ and $s$ are constrained to be non-negative. Using
the auxiliary function method, we can derive an iterative method
to minimize $Q(h,s)$ in order to obtain an estimate for $h$ and
$s$. Let $i$ denote the iteration index, and $h^{i}$ and $s^{i}$
denote the estimates of $h$ and $s$ at the $i$-th iteration, respectively.
The main idea behind the iterative learning method using an auxiliary function is the following lemma from \cite{Lee2000}:
%%%%%%%%%%%%%%%%%%%%%% Lemma %%%%%%%%%%%%%%%%%%%%%%%%%
\begin{lem}
\label{lem:auxiliary}Let $G(h,h^{i})$ be an auxiliary function for the cost function
$L(h)$ such that $G(h,h)=L(h)$ and $G(h,h^{i})\geq L(h)$ for a
given $h^{i}$ and for all $h$. Let $h^{i+1}$ be the new estimate obtained
by minimizing $G(h,h^{i})$ with respect to (w.r.t.) $h$. $L(h)$
is non-increasing under this update, i.e., $L(h^{i+1})\leq L(h^{i})$,
where the equality only holds when $h^{i}$ is a local minimum of $L(h)$. \end{lem}

Considering our problem to minimize $Q(h,s)$, an update rule for $h$ can be derived as follows: assuming that $h^{i}$ and $s^{i}$ are given, $h^{i+1}$  can be computed by minimizing $L(h)=Q(h,s^{i})$ w.r.t. $h$. This is done in two steps: in the first step, an auxiliary function $G(h,h^{i})$ is obtained for $L(h)$. In the second step, the auxiliary function $G(h,h^{i})$ is differentiated w.r.t. $h$, and the derivative is set to zero, leading to a new value for $h$, referred to as $h^{i+1}$, which is a function of $h^{i}$, $s^{i}$, and all the other known parameters. The method is now continued to compute $s^{i+1}$ given $h^{i+1}$ and $s^{i}$. These iterations are executed until a local minimum of the cost function $Q(h,s)$ is obtained. Note that this iterative method can be trivially extended to minimize a function with more than two unknown parameters.

A useful inequality that is often used to obtain an auxiliary function is stated in the following lemma from \cite{Lee2000}:
%%%%%%%%%%%%%%%%%%%%%%% Lemma %%%%%%%%%%%%%%%%%%%%%%%%%
\begin{lem}
\label{lem:jenens}If $\phi(x)$ is a convex function and $a_{\tau}$
are non-negative coefficients for which $\sum_{\tau}a_{\tau}=1$,
Jensen's inequality \cite{Rao1973} can be used to derive the following
inequality:
\begin{equation}
\phi\left(\sum_{\tau}x_{\tau}\right)\leq\sum_{\tau}a_{\tau}\phi\left(\frac{x_{\tau}}{a_{\tau}}\right).\label{eq:jensens_in}
\end{equation}
\end{lem}
\section{Modeling Temporal Dependencies In the Integrated Method USING FRAME STACKING\label{app: Appendix_st}}
Let $\mathbf{y}(t)$ denote the $t-$th column of $\mathbf{Y}$.
We define the $KT_{st}$-dimensional stacked vector $\mathbf{y}_{st}(t)$
as $\mathbf{y}_{st}(t)=[\mathbf{y}^{T}(t)\ldots\mathbf{y}^{T}(t+T_{st}-1)]^{T}$,
where $T$ denotes matrix transpose. The stacked vector $\mathbf{s}_{st}(t)$ is defined
similarly, and the stacked vector $\mathbf{h}_{st}(t)$ is defined as $\mathbf{h}_{st}(t)=[\mathbf{h}^{T}(t)\ldots\mathbf{h}^{T}(t)]^{T}.$
Similarly to (\ref{eq:united_method}), a cost function based on the generalized KL divergence can now be defined as:

{\footnotesize{
\begin{align}
&L_{1,st}=\lambda\sum_{r,t}x\left(r,t\right)+\nonumber\\
&\sum_{l=1}^{T_{st}}\sum_{k,t}KL\left(y_{st}\left(f_l,t\right)\left|\sum_{\tau=0}^{L_h-1}h_{st}\left(f_l,\tau\right)\sum_{r=1}^R w\left(f_l,r\right)x\left(r,t-\tau\right)\right.\right),\label{eq:united_method_st}
\end{align}}}
where $f_l=k+K(l-1)$, and $\mathbf{W}=[w(f_l,r)]$ is a $KT_{st}\times R$-dimensional matrix.
The update rules for $w$ and $x$ remain identical to (\ref{eq:estimation_w_united})
and (\ref{eq:estimation_x_united}), where the update rule for $h$ can be
derived as:
\begin{align}
h^{i+1}\left(k,\tau\right)=&h^{i}\left(k,\tau\right)\times\nonumber\\
&\frac{\sum_{l=1}^{T_{st}}\sum_{t}y_{st}\left(f_l,t\right)\tilde{s}_{st}\left(f_l,t-\tau\right)/\tilde{y}_{st}\left(f_l,t\right)}{\sum_{l=1}^{T_{st}}\sum_{t}\tilde{s}_{st}\left(f_l,t-\tau\right)},\label{eq:update_h_united_stacked}
\end{align}
where $\tilde{s}_{st}(f_l,t)=\sum_{r}w^{i}\left(f_l,r\right)x^{i}\left(r,t\right)$ and $\tilde{y}_{st}\left(f_l,t\right)=\sum_{\tau}h_{st}^{i}\left(f_l,\tau\right)\tilde{s}_{st}\left(f_l,t-\tau\right)$.
By setting $T_{st}=1$, (\ref{eq:estimate_h_united}) is obtained
as a special case of (\ref{eq:update_h_united_stacked}).

After convergence of the iterations, the clean speech spectrogram
is estimated as $\hat{s}(k,t)=G(k,t)y(k,t)$, where the time-varying gain function $G(k,t)$ is obtained
by averaging the overlapping segments, i.e.
\begin{equation}
G\left(k,t\right)=\frac{\sum_{l=1}^{T_{st}}\sum_{r}\hat{w}\left(f_l,r\right)\hat{x}\left(r,t\right)}{\sum_{l=1}^{T_{st}}\sum_{r,\tau}\hat{h}_{st}\left(f_l,\tau\right)\hat{w}\left(f_l,r\right)\hat{x}\left(r,t-\tau\right)},\label{eq:wiener_filtering_stacked}
\end{equation}
where $\hat{(\cdot)}$ is used to denote the obtained estimates after
convergence, and the $KT_{st}$-dimensional vector $\mathbf{h}_{st}^{i}(t)$ is defined
as $\mathbf{h}_{st}^{i}(t)=[\mathbf{h}^{i,T}(t)\ldots\mathbf{h}^{i,T}(t)]^T$. As can be seen, (\ref{eq:wiener_filtering_stacked}) reduces to (\ref{eq:wiener_filtering}) when $T_{st}=1$.
%%%%%%%%%%%%%%%%%%%%%%%%%%%%%%%%%%%%%%%%%%%%%%%%%%%%%%%%%%%%%%%%%%%%%%%%%%%%

\ifCLASSOPTIONcaptionsoff
  \newpage
\fi
%%%%%%%%%%%%%%%%%%%%%%%%%%%%%%%%%%%%%%%%%%%%%%%%%%%%%%%%%%%%%%%%%%%%%%%%%%%%
\bibliographystyle{IEEEtran}
{
%\renewcommand*{\bibfont}{\footnotesize}
%\fontsize{6pt}{8pt plus 0.2pt minus 0.1pt}
C:/Nasser/ASLP2015 N-CTF based Dereverb/NasserRefs
\bibliography{NasserRefs}
}

\end{document}